\documentclass[
  reprint,
  superscriptaddress,
  amsmath,amssymb,
  aps,
  prc,
  floatfix
]{revtex4-2}
\usepackage{slashed}
\usepackage{graphicx}
\graphicspath{{figures/}}
\usepackage{dcolumn}
\usepackage{bm}
\usepackage{booktabs}
\usepackage{tabularx}
\usepackage{color}
\usepackage{hyperref}
\usepackage{ulem}
\usepackage[caption=false,font=normalsize]{subfig}
\usepackage{multirow}
\begin{document}

\title{Finite-temperature quark dynamics in proto-neutron stars: Maximum baryonic mass and radial oscillations}

\author{Xuefeng Ma}
\email[]{maxuefeng21@mails.ucas.ac.cn} 
\affiliation{School of Nuclear Science and Technology, University of Chinese Academy of Sciences, Beijing, 101408, P.R. China} 
\author{Jingdong Shao}
\email[]{shaojingdong19@mails.ucas.ac.cn} 
\affiliation{School of Physical Sciences, University of Chinese Academy of Sciences, Beijing 101408, P.R. China}  
\author{Kun Xu}
\email[]{xukun@bit.edu.cn}
\affiliation{School of Physics, Beijing Institute of Technology, Beijing, 100049,
  P.R. China} 
\author{Mei Huang}
\email[]{huangmei@ucas.ac.cn}
\affiliation{School of Nuclear Science and Technology, University of Chinese Academy of Sciences, Beijing, 101408, P.R. China}

\date{\today}

\begin{abstract}
Finite-temperature quark dynamics can significantly alter proto-neutron star evolution even when the cold equation of state remains unchanged. We quantify this sensitivity by comparing a purely hadronic baseline with two hybrid models incorporating quark matter via the Polyakov–Nambu–Jona-Lasinio (PNJL) framework and its entanglement extension (EPNJL), which isolates the role of finite-temperature chiral–deconfinement coupling. Employing bulk Gibbs constructions for both nuclear liquid–gas and hadron–quark phase coexistence, we compute equilibrium configurations across hot neutrino-trapped, hot neutrino-free, and cold stages, alongside linear radial oscillations of selected hot neutrino-free stars. Introducing quarks lowers the maximum baryonic mass at every stage and reverses its evolutionary ordering: whereas the maximum baryonic mass increases toward the cold limit in the purely hadronic model, it decreases progressively in both hybrid models, opening mass windows for delayed collapse. Crucially, despite sharing the same cold limit, PNJL and EPNJL yield distinct baryonic-mass intervals for mixed-core survival across stages: only EPNJL permits configurations that retain mixed cores across both hot stages, whereas only PNJL allows hot neutrino-free mixed-core stars to possess cold equilibrium counterparts. All tested hot neutrino-free mixed-core stars remain linearly stable, yet their fundamental radial frequencies differ substantially. These findings demonstrate that finite-temperature quark dynamics imprints distinct evolutionary constraints and pulsation signatures on proto-neutron stars beyond cold-matter descriptions.
\end{abstract}

\maketitle

\section{Introduction}
\label{sec:hq-introduction}

A newly formed proto-neutron star is hot and lepton-rich due to neutrino trapping~\cite{Prakash:1996xs}, and it subsequently undergoes neutrino-driven deleptonization and cooling, which alter its thermal state and composition~\cite{Burrows:1986me,Pons:1998mm}. Neglecting accretion and mass ejection, baryonic mass is conserved while gravitational mass decreases through neutrino radiation. Fixed baryonic mass thus provides a natural benchmark for comparing equilibrium configurations across evolutionary stages~\cite{Pons:2001ar} and probing core transitions between hadrons and quarks~\cite{Sagert:2008ka,Hempel:2009vp,Bombaci:2011mx,Klahn:2016uce,Fischer:2017lag}.

Equilibrium configurations depend sensitively on the equation of state (EOS), the entropy per baryon, and the lepton content~\cite{Ghosh:2023tbn,Wu:2025ilj}. While the high-density stiffness of the EOS dictates the baseline maximum mass, a higher entropy per baryon enhances thermal pressure, substantially inflating stellar radii and modifying the core stability limit~\cite{Burgio:2010ek,Raduta:2020fdn}. Concurrently, trapped neutrinos shift the $\beta$-equilibrium condition toward higher proton fractions, which softens the nucleonic symmetry-energy contribution despite adding relativistic lepton pressure, and tends to delay the onset of exotic phases such as deconfined quarks to higher densities~\cite{Prakash:1996xs,Camelio:2017nka,Issifu:2024fuw}. The competition among cooling, deleptonization, and compositional transitions thus governs the structural evolution of proto-neutron stars~\cite{Laskos-Patkos:2026avp}.

The onset and spatial extent of hadron--quark coexistence also depend sensitively on the prevailing thermal and lepton conditions~\cite{Steiner:2000bi,Yasutake:2009kj,Shao:2011nu,Sabatucci:2026qcz}. Generally, finite temperatures favor the higher-degeneracy quark phase and shift deconfinement toward lower baryon densities, whereas neutrino trapping drives hadronic matter closer to isospin symmetry and inhibits deconfinement, pushing the transition threshold to higher densities~\cite{Prakash:1996xs,Sabatucci:2026qcz,Lugones:1997gg,Roark:2018uls}. Beyond shifting the onset density, trapped neutrinos alter the Gibbs phase-equilibrium conditions by introducing electron-lepton number---alongside baryon number and electric charge---as a third conserved charge governed by a finite lepton chemical potential~\cite{Pagliara:2009dg,Hempel:2009vp}. Enforcing global rather than local conservation of charge and lepton number enables the coexisting phases to unevenly partition both electric charge and leptons as the quark fraction increases, substantially broadening the mixed-phase region~\cite{Pagliara:2009dg,Hempel:2009vp,Sabatucci:2026qcz}. Furthermore, as increasing baryon density drives hadron-to-quark conversion along isentropes across the mixed phase, the mismatch in specific entropy between the two phases redistributes thermal energy; consequently, isentropic compression through the coexistence region can exhibit either conventional heating or anomalous cooling with increasing density, feeding back on the local temperature profile and EOS stiffness~\cite{Steiner:2000bi,Roark:2018uls,Salgado:2026xwx}.

Because baryonic mass remains conserved during cooling and deleptonization, the maximum baryonic mass supported at each stage serves as a critical threshold determining whether core-collapse supernova remnants survive as stable neutron stars or undergo delayed collapse into black holes~\cite{Baumgarte:1996xz,Pons:2001ar}. This stability limit depends sensitively on the dense-matter EOS, thermal and lepton profiles, rotation, and the appearance of exotic constituents such as hyperons or deconfined quarks, thereby connecting the microphysics of hot, dense matter to black-hole formation timescales and potentially observable signatures, notably the abrupt termination of neutrino emission~\cite{Keil:1995hw,Pons:2001ar,Fischer:2008rh}. Accordingly, comparisons of equilibrium sequences widely employ gravitational-mass--baryonic-mass ($M_G\text{--}M_B$) relations to assess whether hot configurations possess stable cold counterparts at fixed baryonic mass, or whether shifting maximum-mass limits trigger delayed collapse along the evolutionary path~\cite{Malfatti:2019tpg,Logoteta:2022hvb,Ghaemmaghami:2025atk,Milthaler:2026dhm}.

Complementing static mass-threshold analyses, radial oscillations provide a direct probe of dynamical stability and the response of hot, dense matter~\cite{Chandrasekhar:1964zz,Gondek:1997fd}. In general-relativistic perturbation theory, small spherically symmetric pulsations obey a Sturm-Liouville eigenvalue problem controlled by the background structure and the adiabatic index: dynamical stability requires a positive squared fundamental eigenfrequency ($\omega_0^2 > 0$), with radial instability setting in at $\omega_0^2 = 0$~\cite{Chandrasekhar:1964zz,Kokkotas:2000up}. While $\omega_0 = 0$ coincides with the static maximum-mass turning point along equilibrium barotropic or isentropic sequences, the radial stability and mode spectrum of hybrid stars are further shaped by the phase-conversion kinetics relative to the pulsation timescale~\cite{Pereira:2017rmp,Lugones:2021zsg,Rather:2024hmo}. Consequently, radial modes in cold hybrid stars, hadronic proto-neutron stars, and hot hybrid stars depend sensitively on the EOS, entropy and lepton distributions, and phase-transition softening during cooling and deleptonization~\cite{Sun:2021cez,Sun:2024vdt,Thakur:2025qwl,Rau:2026oxz}.

In this work, we construct the hadronic and hybrid EOSs with a Gibbs construction across three thermodynamic stages, and determine whether an equilibrium configuration of a stage can exist in the later stages, and whether it contains a hadron--quark mixed core. If the mixed core is present, we explore how finite-temperature quark dynamics affects its equilibrium configurations and its persistence in the later stages. We also examine the linear radial stability of representative equilibrium configurations and compare their fundamental frequencies.

The remainder of this paper is organized as follows: Sec.~\ref{sec:model} presents the matter models, thermodynamic stages, construction of EOSs, and expressions for the stellar structure and radial pulsation; Sec.~\ref{sec:level2} presents the resulting phase diagrams, EOSs, and stellar structures, and discusses the persistence of equilibrium configurations across three thermodynamic stages and their radial stability; Sec.~\ref{sec:conclusions} summarizes the main conclusions. The component thermodynamic expressions, phase equilibrium conditions, and detailed background metric and radial pulsation equations are given in the appendices.

\section{Theoretical Models}\label{sec:model}

We use natural units \(\hbar=c=k_B=G=1\) in this paper.

\subsection{Nuclear matter}\label{sec:methodsA}

The nuclear matter is described by a nonlinear relativistic mean-field (RMF) model. Neutrons $\psi_n$ and protons $\psi_p$ interact through scalar $\sigma$, isoscalar-vector $\omega^\mu$, and isovector-vector $\rho_3^\mu$ meson fields~\cite{Routaray:2023spb}.

For uniform matter at rest, we denote the scalar mean field by \(\phi_\sigma\) and the time components of the vector mean fields by \(\omega\equiv\omega_0\) and \(\rho_3\equiv\rho_0^{(3)}\), then the mean-field Lagrangian density is
\begin{equation}
\begin{aligned}
\mathcal L_H^{\mathrm{MF}}={}&
\sum_{i=p,n}\bar\psi_i\Bigl[
i\gamma^\mu\partial_\mu-\gamma^0\left(g_\omega\omega+
\frac{g_\rho}{2}\tau_{3i}\rho_3\right)\\
-&\left(m_N-g_\sigma\phi_\sigma\right)\Bigr]\psi_i+\frac12m_\omega^2\omega^2
+\frac{\zeta}{24}(g_\omega\omega)^4\\
-&U_\sigma(\phi_\sigma)+\frac12m_\rho^2\rho_3^2
+\Lambda_\omega(g_\omega\omega)^2(g_\rho\rho_3)^2,
\end{aligned}
\label{eq:M1}
\end{equation}
where \(\tau_{3p}=+1\), \(\tau_{3n}=-1\), and the scalar potential
\begin{equation}
U_\sigma(\phi_\sigma)=
\frac12m_\sigma^2\phi_\sigma^2
+\frac{\kappa_H}{6}(g_\sigma\phi_\sigma)^3
+\frac{\lambda_H}{24}(g_\sigma\phi_\sigma)^4.
\label{eq:M2}
\end{equation}

Here \(m_N\) is the nucleon mass, and \(m_\sigma\), \(m_\omega\), and \(m_\rho\) are the meson masses. \(g_\sigma\), \(g_\omega\), and \(g_\rho\) are corresponding nucleon--meson coupling constants, and $\kappa_H$, $\lambda_H$, $\Lambda_\omega$ and \(\zeta\) are associated with the interactions of mesons. Table~\ref{tab:rmf} lists the adopted values of these parameters~\cite{Ghosh:2023tbn}. 

Neutrons and protons have same effective mass,
\begin{equation}
M^*=m_N-g_\sigma\phi_\sigma,
\label{eq:M3}
\end{equation}
but different effective chemical potentials
\begin{equation}
\begin{aligned}
\mu_n^*=\mu_n-g_\omega\omega+\frac{g_\rho}{2}\rho_3,\,
\mu_p^*=\mu_p-g_\omega\omega-\frac{g_\rho}{2}\rho_3,
\end{aligned}
\label{eq:M4}
\end{equation}
where \(\mu_n\) and \(\mu_p\) are the chemical potentials of neutrons and protons, respectively. For uniform nuclear matter at \(T>0\), the mean-field grand potential density is
\begin{equation}
\begin{aligned}
\Omega_H={}&U_\sigma(\phi_\sigma)
-\frac12m_\omega^2\omega^2
-\frac{\zeta}{24}(g_\omega\omega)^4-\frac12m_\rho^2\rho_3^2\\
-&\Lambda_\omega(g_\omega\omega)^2(g_\rho\rho_3)^2-\frac{T}{\pi^2}\sum_{i=p,n}
\int_0^\infty dp\,p^2 \times\\
&\Biggl\{
\ln\!\left[1+e^{-\frac{E^*(p)-\mu_i^*}{T}}\right]+\ln\!\left[1+e^{-\frac{E^*(p)+\mu_i^*}{T}}\right]\Biggr\},
\end{aligned}
\label{eq:A1}
\end{equation}
where \(E^*(p)=\sqrt{p^2+M^{*2}}\) is the energy. 

The equilibrium state can be obtained by solving the gap equations:
\begin{equation}
\frac{\partial\Omega_H}{\partial\phi_\sigma}
=\frac{\partial\Omega_H}{\partial\omega}
=\frac{\partial\Omega_H}{\partial\rho_3}=0
\label{eq:A5}
\end{equation}
at fixed \(T\), \(\mu_B\), and \(\mu_{\mathrm{ch}}\). The thermodynamic expressions are given in Appendix~\ref{sec:appA1}, and lepton contributions, stage constraints, and phase equilibrium conditions are specified below.

\begin{table}[!htbp]
\centering
\caption{Parameters of the RMF model.}
\label{tab:rmf}
\begin{tabular}{@{}lr@{}}
\toprule
Parameter & Value \\
\midrule
\(m_N\)          & 939 {MeV}\\
\(m_\sigma\)     & 492.988 {MeV}\\
\(m_\omega\)     & 782.5 {MeV}\\
\(m_\rho\)       & 763 {MeV}\\
\(g_\sigma^2\)   & 97.397 \\
\(g_\omega^2\)   & 160.833 \\
\(g_\rho^2\)     & 201.413 \\
\(\kappa_H\)     & 4.581 {MeV}\\
\(\lambda_H\)    & \(-0.016\) \\
\(\zeta\)        & 0.006 \\
\(\Lambda_\omega\) & 0.045 \\
\bottomrule
\end{tabular}
\end{table}

\subsection{Quark matter}\label{sec:methodsB}

The quark matter is described by the three-flavor Polyakov--Nambu--Jona-Lasinio (PNJL) model and its entanglement extension (EPNJL)~\cite{Ratti:2005jh,Fukushima:2008wg,Sasaki:2011wu}. The Polyakov-loop variables are defined by
\begin{equation}
\Phi=\frac{1}{N_c}\left\langle\operatorname{Tr}_c L\right\rangle,
\qquad
\bar\Phi=\frac{1}{N_c}\left\langle\operatorname{Tr}_c L^\dagger\right\rangle.
\label{eq:M5}
\end{equation}
Here \(N_c=3\) is the number of colors, and \(\operatorname{Tr}_c\) denotes the color trace. The Wilson line in Euclidean time is
\begin{equation}
L(\boldsymbol x)=\mathcal P
\exp\!\left[i\int_0^{1/T}d\tau\,A_4(\boldsymbol x,\tau)\right],
\label{eq:M6}
\end{equation}
where \(A_4\) is the Euclidean temporal component of the background gauge field and \(\mathcal P\) denotes path ordering. We use the logarithmic effective potential of Polyakov loop ~\cite{Roessner:2006xn},
\begin{equation}
\frac{\mathcal U(\Phi,\bar\Phi;T)}{T^4}
=-\frac{a(T)}{2}\Phi\bar\Phi
+b(T)\ln J(\Phi,\bar\Phi),
\label{eq:M13}
\end{equation}
where
\begin{equation}
\begin{aligned}
a(T)&=a_0+a_1\frac{T_0}{T}
+a_2\left(\frac{T_0}{T}\right)^2,\,b(T)=b_3\left(\frac{T_0}{T}\right)^3,\\
J(\Phi,\bar\Phi)&=
1-6\Phi\bar\Phi
+4(\Phi^3+\bar\Phi^3)
-3(\Phi\bar\Phi)^2.
\end{aligned}
\label{eq:M14}
\end{equation}
Here the factor $J>0$ is derived from the \(SU(3)\) Haar measure and $T_0$ denotes the phase transition temperature of Polyakov loop in the absence of quarks. Then the Lagrangian density of the three-flavor PNJL model is
\begin{equation}
\begin{aligned}
\mathcal L_Q={}&
\bar q\left(i\gamma^\mu D_\mu-\hat m_0+\gamma^0\hat\mu\right)q\\
&+G(\Phi,\bar\Phi)\sum_{a=0}^{8}
\left[(\bar q\lambda_a q)^2+
(\bar q i\gamma_5\lambda_a q)^2\right]\\
&-G_D\Bigl\{
\det\nolimits_f[\bar q(1+\gamma_5)q]+\det\nolimits_f[\bar q(1-\gamma_5)q]
\Bigr\}\\
&-\mathcal U(\Phi,\bar\Phi;T).
\end{aligned}
\label{eq:M7}
\end{equation}
with \(q=(u,d,s)^{\mathsf T}\) the quark field in the flavor space, \(\hat m_0=\operatorname{diag}(m_{0u},m_{0d},m_{0s})\) the current quark mass and \(\hat\mu=\operatorname{diag}(\mu_u,\mu_d,\mu_s)\) chemical potential matrices , respectively. The covariant derivative \(D_\mu=\partial_\mu-iA_\mu\) couples quarks to the uniform gauge background, with the gauge coupling absorbed into \(A_\mu\), and the matrices \(\lambda_{1,\ldots,8}\) are the flavor Gell-Mann matrices, with \(\lambda_0=\sqrt{2/3}\,\mathbf1\) and \(\operatorname{tr}(\lambda_a\lambda_b)=2\delta_{ab}\).
The coupling \(G_D\) specifies the six-quark determinant interaction that breaks \(U_A(1)\) symmetry and \(\det_f\) denotes flavor determinant.

The PNJL model uses constant coupling \(G(\Phi,\bar\Phi)=G_S\), whereas the EPNJL model employs a Polyakov-loop-dependent entanglement coupling~\cite{Sakai:2010rp,Sasaki:2011wu},
\begin{equation}
G(\Phi,\bar\Phi)
=G_S\left[1-\alpha_1\Phi\bar\Phi
-\alpha_2(\Phi^3+\bar\Phi^3)\right]
\label{eq:M8}
\end{equation}
Table~\ref{tab:quark} lists the parameters above \cite{Rehberg:1995kh,Sasaki:2011wu,Roessner:2006xn}.

In the mean-field approximation, the constituent quark masses are
\begin{equation}
M_i=m_{0i}-4G(\Phi,\bar\Phi)\varphi_i+2G_D\varphi_j\varphi_k,
\label{eq:M9}
\end{equation}
where \((i,j,k)\) is a cyclic permutation of \((u,d,s)\), and \(\varphi_i=\langle\bar q_i q_i\rangle\) denotes the scalar condensate of quark flavor \(i\). For uniform quark matter at \(T>0\), the mean-field grand potential density is
\begin{equation}
\begin{aligned}
\Omega_Q={}&
2G(\Phi,\bar\Phi)\sum_{i}\varphi_i^2
-4G_D\varphi_u\varphi_d\varphi_s
+\mathcal U(\Phi,\bar\Phi;T)\\
&-2N_c\sum_{i}
\int_{|\boldsymbol p|<\Lambda}
\frac{d^3p}{(2\pi)^3}\,E_i(p)\\
&-2T\sum_{i}\int\frac{d^3p}{(2\pi)^3}
\left[\ln\mathcal F_i^+(p)+\ln\mathcal F_i^-(p)\right]
-\Omega_0,
\end{aligned}
\label{eq:M10}
\end{equation}
where \(E_i(p)=\sqrt{p^2+M_i^2}\), and 
\begin{equation}
\begin{aligned}
\mathcal F_i^+(p)={}&
1+3\Phi e^{-[E_i(p)-\mu_i]/T}\\
&+3\bar\Phi e^{-2[E_i(p)-\mu_i]/T}
+e^{-3[E_i(p)-\mu_i]/T},\\[0.6ex]
\mathcal F_i^-(p)={}&
1+3\bar\Phi e^{-[E_i(p)+\mu_i]/T}\\
&+3\Phi e^{-2[E_i(p)+\mu_i]/T}
+e^{-3[E_i(p)+\mu_i]/T}.
\end{aligned}
\label{eq:M11}
\end{equation}
The constant \(\Omega_0\) is fixed by the vacuum at zero temperature and chemical potential so that the pressure $P_Q=-\Omega_Q$ vanishes at vacuum. The three condensates and two Polyakov-loop variables are determined by the gap equations
\begin{equation}
\frac{\partial\Omega_Q}{\partial\varphi_i}=0,\quad
\frac{\partial\Omega_Q}{\partial\Phi}=0,\quad
\frac{\partial\Omega_Q}{\partial\bar\Phi}=0.
\label{eq:M12}
\end{equation}
Note that the dependence of the entanglement coupling \(G(\Phi,\bar\Phi)\) on Polyakov loop is included in gap equations in the EPNJL model.

\begin{table}[!htbp]
\centering
\caption{Parameters of the PNJL and EPNJL models.}
\label{tab:quark}
\begin{tabular}{@{}lr@{}}
\toprule
Parameter & Value \\
\midrule
\multicolumn{2}{@{}l@{}}{\textbf{Common parameters}} \\
\(\Lambda\)       & 602.3 MeV \\
\(m_{0u}=m_{0d}\)  & 5.5 MeV\\
\(m_{0s}\)        & 140.7 MeV\\
\(G_S\Lambda^2\)  & 1.835 \\
\(G_D\Lambda^5\)  & 12.36 \\
\(T_0\)           & 150 MeV\\
\(a_0\)           & 3.51 \\
\(a_1\)           & \(-2.47\) \\
\(a_2\)           & 15.2 \\
\(b_3\)           & \(-1.75\) \\
\addlinespace
\multicolumn{2}{@{}l@{}}{\textbf{EPNJL entanglement parameters}} \\
\(\alpha_1\)      & 0.25 \\
\(\alpha_2\)      & 0.10 \\
\bottomrule
\end{tabular}
\end{table}

\subsection{Leptons and thermodynamic stages}\label{sec:methodsC}

Leptons are treated as ideal Fermi gases. $\mu_e$ and $\mu_{\nu_e}$ denote the electron and electron neutrino chemical potential, respectively. The lepton chemical potential $\mu_{L_e}$ is defined as the electron neutrino one,
\begin{equation}
    \mu_{\nu_e}=\mu_{L_e},
\end{equation}
and muons are neglected following~\cite{Kunkel:2024otq}. The lepton thermodynamic expressions are given in Appendix~\ref{sec:appA3}. 
Leptons and baryons satisfy the \(\beta\) equilibrium,
\begin{equation}
   \mu_e=-\mu_{\mathrm{ch}}+\mu_{\nu_e},
\end{equation}
where the electric charge chemical potential $\mu_{\mathrm{ch}}$ is associated with the baryon chemical potential $\mu_B$ with definitions
\begin{equation}
    \begin{aligned}
        \mu_B=\mu_n,\quad \mu_{\mathrm{ch}}=\mu_p-\mu_n
    \end{aligned}
\end{equation}
for the nuclear matter, and
\begin{equation}
\begin{aligned}
\mu_u=\frac{\mu_B}{3}+\frac{2\mu_{\mathrm{ch}}}{3},\quad
\mu_d=\mu_s=\frac{\mu_B}{3}-\frac{\mu_{\mathrm{ch}}}{3},
\end{aligned}
\label{eq:M15}
\end{equation}
for the quark matter. For uniform nuclear matter, charge neutrality requires
\begin{equation}
    n_p=n_e^{\mathrm{net}},
\end{equation}
where \(n_e^{\mathrm{net}}=n_{e^-}-n_{e^+}\), and $n_p$ is the net proton density. 

In this work, we consider three representative thermodynamic stages from equilibrium snapshots following Ref.~\cite{Steiner:2000bi}:
\begin{equation}
\begin{aligned}
\mathrm{stage\,1\,(S1)}:&\quad S=1,\qquad Y_{L_e}=0.4,\\
\mathrm{stage\,2\,(S2)}:&\quad S=2,\qquad \mu_{L_e}=0,\\
\mathrm{Cold}:&\quad T=0,\qquad \mu_{L_e}=0.
\end{aligned}
\label{eq:M17}
\end{equation}
The former two represent the hot isentropic stage with the entropy per baryon
\begin{equation}
    S=\frac{s}{n_B}
\end{equation}
given by the ratio of the total entropy density $s$ and the baryon number density \(n_B\). Electron neutrinos are trapped in S1 with a net density \(n_{\nu_e}^{\mathrm{net}}=n_{\nu_e}-n_{\bar\nu_e}\) and contribute to the lepton fraction
\begin{equation}
Y_{L_e}=\frac{n_e^{\mathrm{net}}+n_{\nu_e}^{\mathrm{net}}}{n_B}.
\label{eq:M16}
\end{equation}
In contrast, neutrinos have escaped in and after S2, and the lepton chemical potential $\mu_{L_e}$ reduces to zero.

\subsection{Phase coexistence and Gibbs construction}\label{sec:methodsD}

The hadron--quark (HQ) transition is treated with a bulk Gibbs construction~\cite{Glendenning:1992vb}, and the coexisting phases share the same temperature $T$, baryon chemical potential $\mu_B$, electric chemical potential $\mu_{\mathrm{ch}}$ and pressure $P$,
\begin{equation}
\begin{aligned}
T_H &= T_Q, &
\mu_{B,H} &= \mu_{B,Q}, \\
\mu_{{\mathrm{ch}},H} &= \mu_{{\mathrm{ch}},Q}, &
P_H &= P_Q,
\end{aligned}
\label{eq:M18}
\end{equation}
where the subscripts \(H\) and \(Q\) denote the hadron phase and quark phase, respectively. In S1, the two phases also share the same lepton chemical potential \(\mu_{L_e}\)~\cite{Hempel:2009vp}, and the same lepton pressure cancels from the pressure balance, which thus compares the strong-sector pressures \(P_H\) and \(P_Q\). In the coexisting phases the global charge neutrality is modified by the quark volume fraction $\chi$ and requires
\begin{equation}
(1-\chi)n_{\mathrm{ch},H}
+\chi n_{\mathrm{ch},Q}
-n_e^{\mathrm{net}}=0,
\label{eq:M19}
\end{equation}
where \(n_{\mathrm{ch}}\) is the signed electric charge number density. The interval \(0<\chi<1\) describes the coexistence, while the limits \(\chi=0\) and \(\chi=1\) mark its hadronic and quark boundaries, respectively.

The total baryon number density \(n_B\), entropy density \(s\), energy density \(\varepsilon\), and pressure \(P\) in the coexistence region are
\begin{equation}
\begin{aligned}
n_B&=(1-\chi)n_{\mathrm{B},H}+\chi n_{\mathrm{B},Q},\\
s&=(1-\chi)s_H+\chi s_Q+s_\ell,\\
\varepsilon&=(1-\chi)\varepsilon_H
+\chi\varepsilon_Q+\varepsilon_\ell,\\
P&=P_H+P_\ell=P_Q+P_\ell.
\end{aligned}
\label{eq:M20}
\end{equation}
The subscript \(\ell\) denotes the common lepton contribution for the given stage. Component thermodynamics and the coexistence constraints are given in Appendices~\ref{sec:appA} and~\ref{sec:appB}, respectively.

The same bulk Gibbs construction is applied to the nuclear liquid--gas (LG) transition at low density~\cite{Muller:1995ji}, and the coexisting phases satisfy
\begin{equation}
\begin{aligned}
T_L &= T_G, &
\mu_{B,L} &= \mu_{B,G}, \\
\mu_{\mathrm{ch},L} &= \mu_{\mathrm{ch},G}, &
P_{L} &= P_{G},
\end{aligned}
\end{equation}
here the subscripts \(L\) and \(G\) denote the liquid phase and gas phase, and global charge neutrality is modified by the liquid volume fraction $\eta$
\begin{equation}
(1-\eta)n_{\mathrm{ch},G}
+\eta\, n_{\mathrm{ch},L}
-n_e^{\mathrm{net}}=0.
\label{eq:M21}
\end{equation}
The limits \(\eta=0\) and \(\eta=1\) mark the gas and liquid boundaries of LG coexistence, respectively. The total thermodynamic quantities follow from Eq.~\eqref{eq:M20} by replacing \((H,Q,\chi)\) with \((G,L,\eta)\). 

In this work, we construct two types of EOSs \(P(\varepsilon)\):  one purely hadronic EOS based on the RMF model, and two hybrid EOSs combining the RMF model for hadronic matter with the EPNJL and PNJL models for quark matter, respectively. At each stage, the hadronic EOS crosses the LG coexistence region only and corresponds to the uniform nuclear matter at high density, while the hybrid EOSs cross both the LG and HQ coexistence regions and correspond to the uniform quark matter at high density.

\subsection{Stellar structure and baryonic mass}\label{sec:methodsE}
 
Using the corresponding EOS \(P(\varepsilon)\), we solve the Tolman--Oppenheimer--Volkoff (TOV) equations for hydrostatic equilibrium~\cite{Tolman:1939jz,Oppenheimer:1939ne}
\begin{equation}
\frac{dm}{dr}=4\pi r^2\varepsilon,
\qquad
\frac{dP}{dr}
=-\frac{(\varepsilon+P)(m+4\pi r^3P)}{r(r-2m)},
\label{eq:M22}
\end{equation}
where \(r\) is the radius and \(m(r)\) is the enclosed gravitational mass. For a given central pressure \(P_c\), we integrate outward from \(m(0)=0\) and \(P(0)=P_c\) to the surface pressure cutoff \(P_s\). The radius \(R(P_c)\) is then defined by \(P(R)=P_s\), and the gravitational mass is \(M_G(P_c)=m(R)\). The background metric and the cutoff pressures are given in Appendix~\ref{sec:appC1}.

The total baryon number \(N_B\) enclosed within the radius \(R\) and the corresponding baryonic mass \(M_B\) are defined as~\cite{Rikovska-Stone:2006gml}
\begin{equation}
N_B=4\pi\int_0^R
\frac{n_B(r)r^2}{\sqrt{1-2m(r)/r}}\,dr,
\qquad
M_B=m_bN_B,
\label{eq:M23}
\end{equation}
where the atomic mass constant \(m_{\mathrm u}\simeq931.494\,\mathrm{MeV}\) is adopted as the reference baryon mass \(m_b=m_{\mathrm u}\) \cite{Tiesinga:2021myr}. The equilibrium configurations evolve across stages with conserved \(M_B\), or equivalently \(N_B\).

\subsection{Thresholds and radial stability}\label{sec:methodsF}

To quantitatively determine the quark onset and stability of the equilibrium configurations at fixed $M_B$ across stages, the relevant definitions concerning the baryonic masses are introduced here. In hybrid models, \(P_{\mathrm{on}}^{(i)}\) denotes the total pressure at the hadronic boundary of HQ coexistence at stages $i=\mathrm{S1,\,S2,\,cold}$. \(P_c<P_{\mathrm{on}}^{(i)}\) leads to a hadronic core, while for \(P_c>P_{\mathrm{on}}^{(i)}\) the core composition depends on the central quark volume fraction \(\chi_c=\chi(r=0)\). A mixed core satisfies \(0<\chi_c<1\), while a pure quark core is present with \(\chi_c=1\). \(O_i\) and \(C_i\) denote the baryonic masses at the onset of the mixed core \(P_c=P_{\mathrm{on}}^{(i)}\) and at the gravitational mass peak, respectively:
\begin{equation}
\begin{aligned}
O_i &\equiv M_{B,\mathrm{on}}^{(i)}
    = M_B^{(i)}\!\left(P_{\mathrm{on}}^{(i)}\right),\\
C_i &\equiv M_{B,\max}^{(i)}
    = M_B^{(i)}\!\left(P_{c,\mathrm{peak}}^{(i)}\right),
\end{aligned}
\label{eq:M24}
\end{equation}
where \(P_{c,\mathrm{peak}}^{(i)}\) is the central pressure when \(M_G^{(i)}(P_c)\) peaks. For the purely hadronic EOS, only $C_i$ is defined.

The linear radial stability of the equilibrium configurations, which depends on their fundamental frequencies of the radial modes, is also considered. A displaced fluid element under perturbations follows the equilibrium EOS of the corresponding thermodynamic stage at fixed entropy per baryon \(S\) and under fixed lepton constraints of that stage~\cite{Gondek:1997fd,Sun:2024vdt}. The displacement is written as \(\delta r(r,t)=u(r)e^{i\omega t}\), where \(\omega\) is the angular frequency and \(u(r)\) is the radial amplitude. The variable \(\zeta(r)=r^2e^{-\nu(r)}u(r)\) satisfies the linear radial pulsation equation~\cite{Chandrasekhar:1964zz,Kokkotas:2000up}
\begin{equation}
\frac{d}{dr}\!\left(\Pi\frac{d\zeta}{dr}\right)
+\left(\mathcal{Q}+\omega^2W\right)\zeta=0
\label{eq:M26}
\end{equation}
with the boundary conditions
\begin{equation}
   \lim_{r\rightarrow 0} \zeta\propto r^3,\quad\Delta P(R)=0.
\end{equation}
Here the factor $e^{-\nu(r)}$ in the variable $\zeta$ is associated with the $tt$ component of the static background metric \(g_{tt}=-e^{2\nu(r)}\) given in Appendix~\ref{sec:appC1}. The coefficients \(\Pi\), \(\mathcal{Q}\), and \(W\), together with the Lagrangian pressure perturbation \(\Delta P\), are given in Appendix~\ref{sec:appC2}.

The lowest angular frequency \(\omega_0^2\) corresponds to the fundamental radial mode, whose radial mode has no nodes in \(0<r<R\). A configuration with \(\omega_0^2>0\) is linearly stable against radial perturbations and has a fundamental frequency \(f_0=\omega_0/(2\pi)\). The cases \(\omega_0^2=0\) and \(\omega_0^2<0\) correspond to marginal stability and an exponentially growing radial mode, respectively.

\section{Results}\label{sec:level2}

\subsection{Phase structure, EOSs, and stellar properties}
\label{sec:hq-results-eos-masses}

\begin{figure*}[t]
\centering
\subfloat[][]{%
  \includegraphics[width=0.48\textwidth]{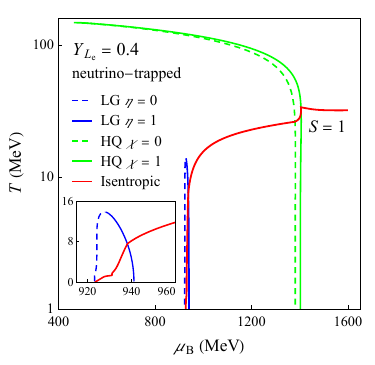}%
}\hfill
\subfloat[][]{%
  \includegraphics[width=0.48\textwidth]{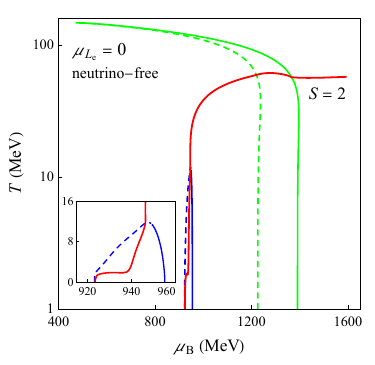}%
}
\caption{The LG (blue) and HQ (green) coexistence boundaries and the corresponding isentropic paths (red) in the EPNJL--RMF hybrid model on the $(\mu_B,T)$ plane. In panel (a), the neutrinos are trapped with a fixed lepton fraction \(Y_{L_e}=0.4\) and the red curve represents the isentropic path in S1 with the entropy per baryon \(S=1\); In panel (b), the neutrinos are free \(\mu_{L_e}=0\) and the red curve represents the isentropic path in S2 with \(S=2\). Dashed (solid) curves mark \(\eta=0\) (\(\eta=1\)) for the LG coexistence and \(\chi=0\) (\(\chi=1\)) for the HQ coexistence.}
\label{fig:phase_epnjl_paths}
\end{figure*}

Figure~\ref{fig:phase_epnjl_paths} shows the LG (blue) and HQ (green) coexistence boundaries in the EPNJL--RMF hybrid model on the \((\mu_B,T)\) plane and the isentropic paths (red). Panel (a) shows a narrow coexistence region in S1 with trapped neutrinos at a fixed lepton fraction \(Y_{L_e}=0.4\) and the corresponding isentropic path with the fixed entropy per baryon \(S=1\), while panel (b) shows a wider coexistence region in S2 with free neutrinos at \(\mu_{L_e}=0\) and the corresponding isentropic path with fixed \(S=2\). 

As the baryon chemical potential $\mu_B$ increases, both paths start from the gas phase boundary of the LG coexistence region (dashed blue) and cross the liquid boundary (solid blue) at $\mu_B\sim940$ MeV and $T\sim10$ MeV. However, the changes in entropy and lepton constraints affect the quark onset within the coexistence region. The path in S2 crosses the hadron boundary (dashed green) of the HQ coexistence region at a higher $T$ and a lower $\mu_B$ than those in S1, indicating a quark onset at lower density but higher temperature in S2.

\begin{figure*}[t]
\centering
\subfloat[][]{%
  \includegraphics[width=0.32\textwidth]{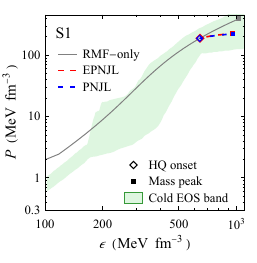}%
}\hfill
\subfloat[][]{%
  \includegraphics[width=0.32\textwidth]{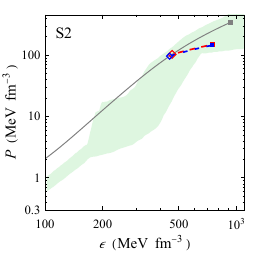}%
}\hfill
\subfloat[][]{%
  \includegraphics[width=0.32\textwidth]{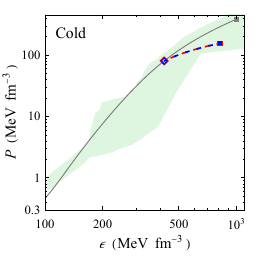}%
}
\caption{EOS $P(\varepsilon)$ in (a) S1, (b) S2, and (c) cold matter compared to the constraint band (cyan) of the cold matter \cite{Annala:2017llu}. The gray solid, red dashed, and blue dashed curves denote the purely hadronic RMF, EPNJL--RMF hybrid, and PNJL--RMF hybrid models, respectively. Open diamonds and filled squares mark the quark onset and the central states of the configurations with peak gravitational masses.}
\label{fig:hq-eos-stages}
\end{figure*}

Figure~\ref{fig:hq-eos-stages} compares EOSs $P(\varepsilon)$ of the purely hadronic RMF model (gray), EPNJL--RMF hybrid model (red) and PNJL-RMF hybrid model (blue) in S1 (panel a), S2 (panel b), and cold matter (panel c) to the constraints band (cyan) for the cold matter~\cite{Annala:2017llu,Zhang:2022uin}. The purely hadronic EOS only deviates from the hybrid ones at energy densities larger than that of the quark onset, and at given $\varepsilon$ in this range, the hybrid models show lower pressure and thus softer EOSs due to the quark onset, compared to that of the purely hadronic model. Within the hybrid models, the PNJL and EPNJL models show similar EOSs.

\begin{figure*}[t]
\centering
\subfloat[][]{
  \includegraphics[width=0.32\textwidth]{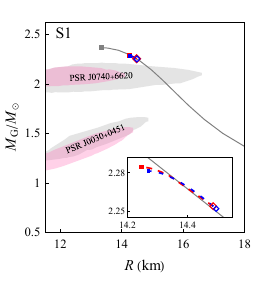}
}\hfill
\subfloat[][]{
  \includegraphics[width=0.32\textwidth]{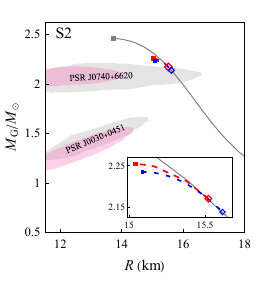}
}\hfill
\subfloat[][]{
  \includegraphics[width=0.32\textwidth]{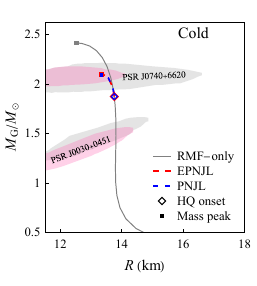}
}
\caption{Gravitational masses $M_G$ versus radius $R$ in the purely hadronic model (gray), EPNJL--RMF hybrid model (dashed red) and PNJL--RMF hybrid model (dashed blue), in (a) S1, (b) S2, and (c) cold matter. Open diamonds and filled squares mark the quark onset and the peak gravitational mass, respectively. The insets highlight the quark onsets and peak gravitational masses of the hybrid models in S1 and S2. The pink and gray shaded regions show the 90\% credible mass--radius regions of two observation events PSR~J0030+0451~\cite{Riley:2019yda,Miller:2019cac} and PSR~J0740+6620~\cite{Riley:2021pdl,Miller:2021qha}.}
\label{fig:mr_three_stage_models}
\end{figure*}

Figure~\ref{fig:mr_three_stage_models} shows the corresponding mass--radius relations in S1 (panel a), S2 (panel b) and cold matter (panel c) compared to two observation events PSR~J0030+0451 \cite{Riley:2019yda,Miller:2019cac} and PSR~J0740+6620 \cite{Riley:2021pdl,Miller:2021qha}. Beyond the quark onset mass, the EPNJL--RMF hybrid model (dashed red) and PNJL--RMF hybrid model (dashed blue) depart from the purely hadronic model (gray) and reach lower peak gravitational masses $M_G\sim2.1M_\odot$ than that in the purely hadronic model \(2.41\,M_\odot\) due to the softer EOSs, while within the hybrid models, the PNJL and EPNJL model have similar mass--radius relations with a slight difference of the peak $M_G$ as the inset shows, as expected from the similar EOSs.

\subsection{Maximum baryonic masses and mixed-core onset thresholds}
\label{sec:hq-results-capacity-onset}

\begin{figure}[t]
\centering
\includegraphics[width=\columnwidth]{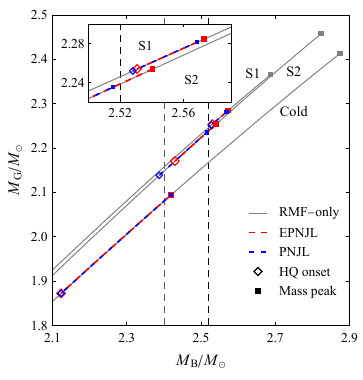}
\caption{Gravitational mass \(M_G\) versus baryonic mass \(M_B\) in the purely hadronic model (gray), EPNJL--RMF hybrid model (dashed red) and PNJL--RMF hybrid model (dashed blue) in the three stages. Open diamonds and filled squares mark the baryonic masses at the quark onset \(M_{B,\mathrm{on}}\) and the peak gravitational mass \(M_{B,\max}\), respectively. Vertical black dashed lines mark the reference masses \(M_B/M_\odot=2.40\) and \(2.52\) and the inset highlights the region where $M_G$ of hybrid models peaks in S1 and S2.}
\label{fig:hq-mgmb-stages}
\end{figure}

Figure~\ref{fig:hq-mgmb-stages} shows the \(M_G(M_B)\) relations in the purely hadronic model (gray), EPNJL--RMF hybrid model (dashed red) and PNJL--RMF hybrid model (dashed blue) in the three stages. In a given stage, the hybrid models give lower maximum baryonic masses $C_i=M^i_{B,{\mathrm{max}}}$ than that in the purely hadronic model. Within the hybrid models, although the PNJL and EPNJL share the similar mass-radius relations, they give different mixed-core onset masses $O_i$ and maximum baryonic masses $C_i$ in the two hot stages that lead to different equilibrium configurations and core compositions.

{\setlength{\tabcolsep}{12pt}
\begin{table*}[tbp]
\centering
\caption{Mixed-core onset masses, maximum baryonic masses, and core compositions at \(M_{B,\max}\) in different models and stages.}
\label{tab:mass-thresholds}
\begin{tabular}{clrrc}\toprule
Model & Stage & $O_i/M_\odot$ & $C_i/M_\odot$
& Core \\\midrule

\multirow{3}{*}{\shortstack{Purely\\hadronic}} & S1
& \multicolumn{1}{c}{--} & 2.68730 & hadron \\
 & S2
& \multicolumn{1}{c}{--} & 2.82369 & hadron \\
 & Cold
& \multicolumn{1}{c}{--} & 2.87418 & hadron \\[3pt]\cline{1-5}

\rule{0pt}{3ex}\multirow{3}{*}{\shortstack{EPNJL--RMF\\hybrid}} & S1
& 2.53068 & 2.57286 & mixed \\
 & S2
& 2.42969 & 2.54033 & mixed \\
  & Cold
& 2.12330 & 2.41919 & mixed \\[3pt]\cline{1-5}
\rule{0pt}{3ex}\multirow{3}{*}{\shortstack{PNJL--RMF\\hybrid}} & S1
& 2.52799 & 2.56866 & mixed \\
 & S2
& 2.38756 & 2.51538 & mixed \\ 
 & Cold & 2.12330 & 2.41919 & mixed \\
\bottomrule
\end{tabular}
\end{table*}}

The equilibrium configurations are compared with conserved $M_B$ across different stages. Since \(C_i\) sets the upper limit of the baryonic mass in each stage, the equilibrium counterpart of an equilibrium configuration does not exist in later stages if $M_B$ lies beyond $C_i$ of the later stages. In particular, the comparison between \(O_i\) and $C_i$ determines whether a mixed-core can exist. Table~\ref{tab:mass-thresholds} summarizes the mixed-core onset masses \(O_i\), maximum baryonic masses \(C_i\) and the core compositions at \(M_{B,\max}\) in different models in S1, S2, and the cold stage. In the purely hadronic model, $C_i$ increases from S1 through S2 to the cold stage,
\(C_{\mathrm{S1}}<C_{\mathrm{S2}}<C_{\mathrm{Cold}}\),
whereas the hybrid models both give the opposite trends
\(C_{\mathrm{Cold}}<C_{\mathrm{S2}}<C_{\mathrm{S1}}\),
and reach the largest difference \(0.455\,M_\odot\) in the cold stage. Meanwhile the hybrid models also give decreasing $O_i$,
\(O_{\mathrm{Cold}}<O_{\mathrm{S2}}<O_{\mathrm{S1}}\)
with \(O_i<C_i\) in each stage, and thus indicate the possibilities of the existence of the mixed cores in the given stage. However, the existence of the mixed cores in each stage does not guarantee that they survive across stages, and whether the equilibrium counterparts of these mixed-core configuration exist in later stages depends on the comparison between $O_i$ in hot stages and $C_i$ in later stages.

\begin{table*}[tbp]
\centering
\caption{Core compositions and existence of equilibrium counterparts in different intervals of $M_B/M_\odot$ in different stages. The label \textquotedblleft{}absent\textquotedblright{} means the absence of the equilibrium counterparts.}
\label{tab:hq-stage-classifications}
\setlength{\tabcolsep}{12pt}
\begin{tabular}{ccccc}
\toprule
 & & \multicolumn{3}{c}{Core compositions}\\
\raisebox{5pt}{Model} & \raisebox{5pt}{Intervals of $M_B/M_\odot$} & S1 & S2 & Cold \\
\midrule
\multirow{5}{*}{\shortstack{EPNJL--RMF\\hybrid}} & $(2.54033,\,2.57286)$ & mixed & absent & absent \\
 & $(2.53068,\,2.54033)$ & mixed & mixed & absent \\
 & $(2.42969,\,2.53068)$ & hadron & mixed & absent \\
 & $(2.41919,\,2.42969)$ & hadron & hadron & absent \\
 & $(2.12330,\,2.41919)$ & hadron & hadron & mixed \\
\midrule
\multirow{5}{*}{\shortstack{PNJL--RMF\\hybrid}} & $(2.52799,\,2.56866)$ & mixed & absent & absent \\
 & $(2.51538,\,2.52799)$ & hadron & absent & absent \\
 & $(2.41919,\,2.51538)$ & hadron & mixed & absent \\
 & $(2.38756,\,2.41919)$ & hadron & mixed & mixed \\
 & $(2.12330,\,2.38756)$ & hadron & hadron & mixed \\
\bottomrule
\end{tabular}
\end{table*}

Table~\ref{tab:hq-stage-classifications} lists the classifications of the core compositions in different intervals of $M_B/M_\odot$ in different stages. The label ``absent" means the absence of the equilibrium counterparts. In both hybrid models, although a mixed-core configuration can exist in hot stages,  no equilibrium counterpart in the cold stage exists with $M_B>2.42M_\odot$ regardless of the core compositions in hot stages. In contrast, the equilibrium counterparts in the interval $2.12<M_B/M_\odot<2.42$ convert to a mixed-core configuration in the cold stage from a hadron-core one in hot stages, while only hadronic configurations exist with $M_B$ below $2.12M_\odot$.

Moreover, several cases exist only in one of the hybrid models. In the EPNJL--RMF model, there exist $M_B$ intervals where the core compositions of S1 and S2 are the same and no equilibrium counterpart exists in the cold stages; equilibrium counterparts of all the mixed-core configurations in hot stages are absent in the cold stage. In contrast, in the PNJL--RMF model, some S2 mixed-core configurations possess cold equilibrium counterparts: an S1 hadron-core configuration can either convert into an S2 mixed-core star that survives to the cold stage as a mixed-core star (e.g., at $M_B = 2.40\,M_\odot$ in Fig.~\ref{fig:hq-mgmb-stages}), or lose its equilibrium counterpart altogether upon entering S2 while EPNJL still supports a mixed-core star (e.g., at $M_B = 2.52\,M_\odot$ in Fig.~\ref{fig:hq-mgmb-stages}). These distinguishing cases in each model can serve as a probe for different quark descriptions.

\subsection{Radial oscillations and stability}
\label{sec:hq-results-radial}

\begin{figure}[tbp]
\centering
\includegraphics[width=\columnwidth]{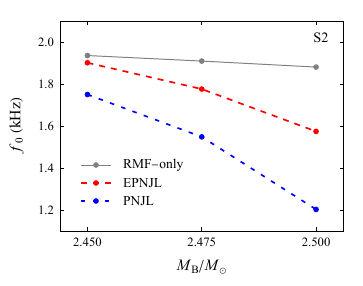}
\caption{Fundamental radial frequencies of purely hadronic (gray), EPNJL--RMF hybrid (red), and PNJL--RMF (blue) configurations at three baryonic masses in S2.}
\label{fig:hq-radial-three-models}
\end{figure}

Figure~\ref{fig:hq-radial-three-models} shows the fundamental frequencies of purely hadronic (gray), EPNJL--RMF hybrid (red), and PNJL--RMF (blue) configurations at three baryonic masses \(M_B/M_\odot=2.450\), \(2.475\), and \(2.500\). All the nine configurations are radially stable with \(\omega_0^2>0\) and satisfy
\begin{equation}
f_0^{\text{RMF-only}}>
f_0^{\mathrm{EPNJL}}>
f_0^{\mathrm{PNJL}}
\label{eq:hq-radial-frequency-ordering}
\end{equation}
in S2. The frequency separations increase with $M_B$. At \(M_B=2.500\,M_\odot\), the fundamental frequencies are approximately \(1.881\), \(1.574\), and \(1.203\,\mathrm{kHz}\), respectively. Relative to the purely hadronic one, the frequencies are lower by approximately \(16.3\%\) for the EPNJL--RMF hybrid model and \(36.0\%\) for the PNJL--RMF hybrid model. Therefore, the fundamental frequency can distinguish the EPNJL and PNJL configurations even when both are radially stable and share the same baryonic mass and core composition.

Here all the six configurations with mixed-cores in the hybrid models are linearly stable. However, none has a cold equilibrium counterpart, because $M_B$ exceeds the cold maximum baryonic mass \(C_{\mathrm{Cold}}\simeq2.41919\,M_\odot\). Linear radial stability at a hot stage therefore does not guarantee a cold equilibrium counterpart at fixed baryonic mass.

\section{Conclusions}
\label{sec:conclusions}

We have investigated how finite-temperature quark dynamics shapes the structural evolution and radial stability of proto-neutron stars by comparing a purely hadronic RMF baseline with hybrid models based on the PNJL and EPNJL quark descriptions. Because both hybrid schemes share the same hadronic sector, their differences across the hot neutrino-trapped (S1) and hot neutrino-free (S2) stages directly isolate the impact of finite-temperature chiral--deconfinement entanglement. Employing Gibbs constructions at conserved baryonic mass, we find that introducing quarks lowers both the peak gravitational mass and the maximum baryonic mass at every stage, with all hybrid maximum-mass configurations hosting hadron--quark mixed cores. Crucially, quark softening reverses the evolutionary ordering of the maximum baryonic mass---changing it from increasing toward the cold limit in the purely hadronic baseline to progressively decreasing in both hybrid models. Consequently, hot hybrid configurations within a baryonic-mass window of width ${\sim}0.1\,M_\odot$ above the cold threshold possess no cold equilibrium counterparts, providing a static basis for identifying massive proto-neutron star candidates prone to delayed black-hole collapse.

Despite sharing an identical cold equation of state and yielding nearly indistinguishable hot mass--radius curves, the PNJL and EPNJL models produce qualitatively different stage classifications at fixed baryonic mass. Specifically, only EPNJL permits mixed-core configurations to persist across both hot stages (S1 and S2), whereas only PNJL allows hot neutrino-free (S2) mixed-core stars to survive to the cold stage as stable hybrid stars. Future work should examine how robustly these threshold orderings hold against variations in hadronic and quark interaction parameters, as well as the inclusion of rapid uniform or differential rotation, which can substantially elevate maximum-mass limits and shift the delayed-collapse windows~\cite{Breu:2016ufb,Baumgarte:1999cq}.

Finally, linear radial oscillation analyses under stage-equilibrium isentropic perturbations show that all nine tested S2 configurations at $M_B/M_\odot = 2.450$, $2.475$, and $2.500$ satisfy $\omega_0^2 > 0$ and remain dynamically stable, even though none of the six mixed-core stars possesses a cold equilibrium counterpart. Nevertheless, their fundamental frequencies clearly separate the models according to $f_0^{\text{RMF-only}} > f_0^{\text{EPNJL}} > f_0^{\text{PNJL}}$: at $M_B = 2.500\,M_\odot$, the EPNJL and PNJL frequencies drop by approximately $16.3\%$ and $36.0\%$ (several hundred hertz) relative to the hadronic baseline, respectively. Fundamental radial modes thus offer a sensitive dynamical signature that distinguishes finite-temperature quark interactions beyond static core-composition classifications. Extending the perturbation treatment to frozen-composition responses~\cite{Ghosh:2024auj} and finite phase-conversion rates across extended mixed phases remains an important direction for future study.

\begin{acknowledgments}
This work is supported in part by the National Natural Science 
Foundation of China (NSFC) Grant Nos: 12235016, 12221005, 12505148.
\end{acknowledgments}

\appendix
\renewcommand{\thesection}{\Alph{section}}
\renewcommand{\thesubsection}{\thesection.\arabic{subsection}}
\makeatletter
\renewcommand{\p@subsection}{}
\makeatother
\section{Thermodynamical quantities}\label{sec:appA}

\subsection{Nuclear Matter}\label{sec:appA1}

The thermodynamic expressions are given by the distribution function. For the nuclear matter with finite temperature $T$, distributions of the nucleons and antinucleons are
\begin{equation}
f_i(p)=\frac{1}{e^{[E^*(p)-\mu_i^*]/T}+1},
\qquad
\bar f_i(p)=\frac{1}{e^{[E^*(p)+\mu_i^*]/T}+1},
\label{eq:A3}
\end{equation}
with $i=p,n$, and for the cold matter  \(T\to0\), the distributions reduce to
\begin{equation}
\begin{aligned}
f_i(p)&\longrightarrow
\Theta\!\left(\mu_i^*-E^*(p)\right),\\
\bar f_i(p)&\longrightarrow
\Theta\!\left(-\mu_i^*-E^*(p)\right),
\end{aligned}
\label{eq:A9}
\end{equation}
where \(\Theta\) is the Heaviside step function. Then the number and scalar densities are
\begin{equation}
\begin{aligned}
n_i&=\frac{1}{\pi^2}\int_0^\infty \mathrm{d}p\,p^2
\left[f_i(p)-\bar f_i(p)\right],\\
n_{s,i}&=\frac{1}{\pi^2}\int_0^\infty \mathrm{d}p\,p^2
\frac{M^*}{E^*(p)}
\left[f_i(p)+\bar f_i(p)\right]
\end{aligned}
\label{eq:A4}
\end{equation}
and the baryon and signed charge number densities are
\begin{equation}
\begin{aligned}
n_{B,H}&=-\left(\frac{\partial\Omega_H}{\partial\mu_B}\right)_{T,\mu_{\mathrm{ch}}}
=n_n+n_p,\\
n_{\mathrm{ch},H}&=-\left(\frac{\partial\Omega_H}{\partial\mu_{\mathrm{ch}}}\right)_{T,\mu_B}
=n_p.
\end{aligned}
\label{eq:A10}
\end{equation}
The kinetic pressure of nucleons is
\begin{equation}
P_i^{\mathrm{kin}}=
\frac{1}{3\pi^2}\int_0^\infty dp\,
\frac{p^4}{E^*(p)}\left[f_i(p)+\bar f_i(p)\right],
\label{eq:A11}
\end{equation}
and the total pressure of the strong sector is
\begin{equation}
\begin{aligned}
P_H={}&\sum_{i=p,n}P_i^{\mathrm{kin}}
-U_\sigma(\phi_\sigma)
+\frac12m_\omega^2\omega^2
+\frac{\zeta}{24}(g_\omega\omega)^4\\
&+\frac12m_\rho^2\rho_3^2
+\Lambda_\omega(g_\omega\omega)^2(g_\rho\rho_3)^2.
\end{aligned}
\label{eq:A12}
\end{equation}
The entropy density is
\begin{equation}
\begin{aligned}
s_H
&=-\left(\frac{\partial\Omega_H}{\partial T}\right)_{\mu_B,\mu_{\mathrm{ch}}}\\
&=-\frac{1}{\pi^2}\sum_{i=p,n}
\int_0^\infty dp\,p^2\\
&\quad\times\Bigl[
f_i\ln f_i+(1-f_i)\ln(1-f_i)\\
&\qquad+
\bar f_i\ln\bar f_i+(1-\bar f_i)\ln(1-\bar f_i)
\Bigr],
\end{aligned}
\label{eq:A13}
\end{equation}
and the energy density reads
\begin{equation}
\varepsilon_H
=T s_H+\mu_B n_{B,H}
+\mu_{\mathrm{ch}} n_{\mathrm{ch},H}-P_H.
\label{eq:A14}
\end{equation}
With the kinetic energy density of nucleons
\begin{equation}
\varepsilon_i^{\mathrm{kin}}=
\frac{1}{\pi^2}\int_0^\infty dp\,p^2E^*(p)
\left[f_i(p)+\bar f_i(p)\right]
\label{eq:A15}
\end{equation}
and the energy density can be explicitly written as
\begin{equation}
\begin{aligned}
\varepsilon_H={}&\sum_{i=p,n}\varepsilon_i^{\mathrm{kin}}
+U_\sigma(\phi_\sigma)
+\frac12m_\omega^2\omega^2
+\frac{\zeta}{8}(g_\omega\omega)^4\\
&+\frac12m_\rho^2\rho_3^2
+3\Lambda_\omega(g_\omega\omega)^2(g_\rho\rho_3)^2.
\end{aligned}
\label{eq:A16}
\end{equation}

\subsection{Quark Matter}\label{sec:appA2}

For brevity we first define
\begin{equation}
\begin{aligned}
z_i^\pm(p)&=\exp\!\left[-\frac{E_i(p)\mp\mu_i}{T}\right],\\
\mathcal F_i^+&=1+3\Phi z_i^+
+3\bar\Phi(z_i^+)^2+(z_i^+)^3,\\
\mathcal F_i^-&=1+3\bar\Phi z_i^-
+3\Phi(z_i^-)^2+(z_i^-)^3,
\end{aligned}
\label{eq:A17}
\end{equation}
where $ i=u,d,s$ and the quark and antiquark distributions are
\begin{equation}
\begin{aligned}
f_i^+&=\frac{\Phi z_i^+
+2\bar\Phi(z_i^+)^2+(z_i^+)^3}{\mathcal F_i^+},\\
f_i^-&=\frac{\bar\Phi z_i^-
+2\Phi(z_i^-)^2+(z_i^-)^3}{\mathcal F_i^-}.
\end{aligned}
\label{eq:A18}
\end{equation}
The three scalar condensates are determined by
\begin{equation}
\begin{aligned}
\varphi_i={}&-\frac{N_c}{\pi^2}
\int_0^\Lambda dp\,p^2\frac{M_i}{E_i(p)}\\
&+\frac{N_c}{\pi^2}
\int_0^\infty dp\,p^2\frac{M_i}{E_i(p)}
\left(f_i^++f_i^-\right),
\end{aligned}
\label{eq:A19}
\end{equation}
and the two Polyakov-loop variables satisfy
\begin{equation}
\begin{aligned}
0={}&\frac{\partial\mathcal U}{\partial\Phi}
+2G_S\left(\alpha_1\bar\Phi+3\alpha_2\Phi^2\right)\sum_i\varphi_i^2\\
&-\frac{3T}{\pi^2}\sum_i\int_0^\infty dp\,p^2
\left[\frac{z_i^+}{\mathcal F_i^+}
+\frac{(z_i^-)^2}{\mathcal F_i^-}\right],\\[0.6ex]
0={}&\frac{\partial\mathcal U}{\partial\bar\Phi}
+2G_S\left(\alpha_1\Phi+3\alpha_2\bar\Phi^2\right)\sum_i\varphi_i^2\\
&-\frac{3T}{\pi^2}\sum_i\int_0^\infty dp\,p^2
\left[\frac{(z_i^+)^2}{\mathcal F_i^+}
+\frac{z_i^-}{\mathcal F_i^-}\right],
\end{aligned}
\label{eq:A20}
\end{equation}
where $(\alpha_1,\alpha_2)=(0.25,0.1)$ in the EPNJL model and \(\alpha_1=\alpha_2=0\) in the PNJL model. The derivative of the logarithmic Polyakov potential can be explicitly written as
\begin{equation}
\begin{aligned}
\frac{\partial\mathcal U}{\partial\Phi}
&=T^4\left[-\frac{a(T)}{2}\bar\Phi
+b(T)\frac{J_\Phi}{J}\right],\\
\frac{\partial\mathcal U}{\partial\bar\Phi}
&=T^4\left[-\frac{a(T)}{2}\Phi
+b(T)\frac{J_{\bar\Phi}}{J}\right],\\
J_\Phi&\equiv\frac{\partial J}{\partial\Phi}=-6\bar\Phi+12\Phi^2-6\Phi\bar\Phi^2,\\
J_{\bar\Phi}&\equiv \frac{\partial J}{\partial\bar\Phi} =-6\Phi+12\bar\Phi^2-6\Phi^2\bar\Phi.
\end{aligned}
\label{eq:A22}
\end{equation}

The number density of each quark flavor is
\begin{equation}
n_i=\frac{N_c}{\pi^2}\int_0^\infty dp\,p^2
\left(f_i^+-f_i^-\right),
\label{eq:A23}
\end{equation}
and the baryon and signed charge number densities are therefore
\begin{equation}
n_{B,Q}=\frac{n_u+n_d+n_s}{3},
\qquad
n_{\mathrm{ch},Q}=\frac{2n_u-n_d-n_s}{3}.
\label{eq:A24}
\end{equation}
The entropy density is
\begin{equation}
\begin{aligned}
s_Q={}&-\left(\frac{\partial\mathcal U}{\partial T}\right)_{\Phi,\bar\Phi}+\frac{1}{\pi^2}\sum_i\int_0^\infty dp\,p^2 \times\\
&\Biggl[
\ln\mathcal F_i^++\ln\mathcal F_i^-+\frac{3}{T}\Bigl((E_i-\mu_i)f_i^+
+(E_i+\mu_i)f_i^-\Bigr)\Biggr],
\end{aligned}
\label{eq:A26}
\end{equation}
where the first term is given by the temperature-dependent Polyakov loop potential
\begin{equation}
\begin{aligned}
\left(\frac{\partial\mathcal U}{\partial T}\right)_{\Phi,\bar\Phi}
={}&-\frac12\left(4a_0T^3+3a_1T_0T^2
+2a_2T_0^2T\right)\Phi\bar\Phi\\
&+b_3T_0^3\ln J.
\end{aligned}
\label{eq:A27}
\end{equation}
The pressure and energy density satisfy
\begin{equation}
P_Q=-\Omega_Q,\,\varepsilon_Q=T s_Q+\mu_B n_{B,Q}
+\mu_{\mathrm{ch}} n_{\mathrm{ch},Q}-P_Q,
\label{eq:A25}
\end{equation}
and the energy density can also be written as
\begin{equation}
\begin{aligned}
\varepsilon_Q={}&2G\sum_i\varphi_i^2
-4G_D\varphi_u\varphi_d\varphi_s
-\frac{N_c}{\pi^2}\sum_i\int_0^\Lambda dp\,p^2E_i(p)\\
&+\frac{N_c}{\pi^2}\sum_i\int_0^\infty dp\,p^2E_i(p)
\left(f_i^++f_i^-\right)\\
&+\mathcal U
-T\left(\frac{\partial\mathcal U}{\partial T}\right)_{\Phi,\bar\Phi}
-\Omega_0.
\end{aligned}
\label{eq:A28}
\end{equation}
For cold quark matter $T=0$, the distributions also reduce to the Heaviside step functions, and the zero temperature limit leads to \(G=G_S\), \(\mathcal U=0\), and \(s_Q=0\).

\subsection{Lepton}\label{sec:appA3}

For an ideal electron–positron gas, the distributions are
\begin{equation}
\begin{aligned}
f_{e^-}(p)&=\frac{1}{e^{[E_e(p)-\mu_e]/T}+1},\\
f_{e^+}(p)&=\frac{1}{e^{[E_e(p)+\mu_e]/T}+1},
\end{aligned}
\label{eq:A29}
\end{equation}
where $E_e(p)=\sqrt{p^2+m_e^2}$ is the energy and \(m_e\) is the electron mass.
The number density, pressure, energy density and entropy density are
\begin{equation}
\begin{aligned}
n_e^{\mathrm{net}}
&=\frac{1}{\pi^2}\int_0^\infty dp\,p^2
\left(f_{e^-}-f_{e^+}\right),\\
P_e&=\frac{1}{3\pi^2}\int_0^\infty dp\,
\frac{p^4}{E_e(p)}
\left(f_{e^-}+f_{e^+}\right),\\
\varepsilon_e
&=\frac{1}{\pi^2}\int_0^\infty dp\,p^2E_e(p)
\left(f_{e^-}+f_{e^+}\right),\\
s_e&=\frac{\varepsilon_e+P_e-\mu_e n_e^{\mathrm{net}}}{T}.
\end{aligned}
\label{eq:A30}
\end{equation}
In cold matter $T=0$, the distributions also reduce to the Heaviside step functions, and the entropy density reduces to zero \(s_e=0\).

Similarly, the neutrino distributions are
\begin{equation}
f_{\nu_e}(p)=\frac{1}{e^{(p-\mu_{\nu_e})/T}+1},
\qquad
f_{\bar\nu_e}(p)=\frac{1}{e^{(p+\mu_{\nu_e})/T}+1},
\label{eq:A32}
\end{equation}
where the small rest mass of neutrinos are neglected $m_{\nu_e}\approx0$.
Then the number density, pressure, energy density, and entropy density are
\begin{equation}
\begin{aligned}
n_{\nu_e}^{\mathrm{net}}
&=\frac{\mu_{\nu_e}^3}{6\pi^2}
+\frac{\mu_{\nu_e}T^2}{6},\\
P_{\nu_e}
&=\frac{\mu_{\nu_e}^4}{24\pi^2}
+\frac{\mu_{\nu_e}^2T^2}{12}
+\frac{7\pi^2T^4}{360},\\
\varepsilon_{\nu_e}&=3P_{\nu_e},\\
s_{\nu_e}
&=\frac{\mu_{\nu_e}^2T}{6}
+\frac{7\pi^2T^3}{90}.
\end{aligned}
\label{eq:A33}
\end{equation}

The total lepton contributions are
\begin{equation}
\begin{aligned}
(P_\ell,\varepsilon_\ell,s_\ell)_{\mathrm{S1}}
&=(P_e+P_{\nu_e},\,
\varepsilon_e+\varepsilon_{\nu_e},\,
s_e+s_{\nu_e}),\\
(P_\ell,\varepsilon_\ell,s_\ell)_{\mathrm{S2,Cold}}
&=(P_e,\varepsilon_e,s_e).
\end{aligned}
\label{eq:A34}
\end{equation}
In S1, both electrons and neutrinos contribute under the constraint \(Y_{L_e}=0.4\), while neutrinos are free to escape in S2 and in the cold matter.

\section{Phase equilibrium and EOS construction}\label{sec:appB}

For each phase at the common \((T,\mu_B,\mu_{\mathrm{ch}})\) in the HQ and LG coexistence regions, the Gibbs construction satisfies the constraints on the given baryon number density, charge number density, lepton number density and entropy density.

For the given baryon number density \(n_\star\), entropy per baryon \(S_\star\) and lepton fraction \(Y_\star\), the general constraints are
\begin{equation}
\begin{cases}
P_a-P_b=0,\\[1mm]
(1-v)n_{B,a}+v n_{B,b}=n_\star,\\[1mm]
(1-v)n_{\mathrm{ch},a}+v n_{\mathrm{ch},b}
-n_e^{\mathrm{net}}=0,\\[1mm]
(1-v)s_a+v s_b+s_\ell=S_\star n_\star,\\[1mm]
n_e^{\mathrm{net}}+n_{\nu_e}^{\mathrm{net}}
=Y_\star n_\star.
\end{cases}
\label{eq:B1}
\end{equation}
Here the label \((a,b,v)\) corresponds to $(G,L,\eta)$ for the LG coexistence and to $(H,Q,\chi)$ for the HQ coexistence. In S1, \(S_\star=1\) and \(Y_\star=0.4\) are fixed, while in S2, \(S_\star=2\) is fixed and neutrino contributions vanish. For the cold matter \(T=0\), the entropy constraint also vanishes. 

\section{Background metric and radial oscillations}\label{sec:appC}

\subsection{Background metric}\label{sec:appC1}

The static, spherically symmetric metric is
\begin{equation}
\begin{aligned}
ds^2&=-e^{2\nu(r)}dt^2+e^{2\lambda(r)}dr^2
+r^2\left(d\theta^2+\sin^2\theta\,d\phi^2\right),\\
e^{-2\lambda(r)}&=1-\frac{2m(r)}{r},
\end{aligned}
\label{eq:C1}
\end{equation}
where \(m(r)\) is the enclosed gravitational mass. The temporal component satisfies
\begin{equation}
\frac{d\nu}{dr}
=-\frac{1}{\varepsilon+P}\frac{dP}{dr}
=\frac{m+4\pi r^3P}{r(r-2m)},
\label{eq:C2}
\end{equation}
and its solution
\begin{equation}
e^{2\nu(r)}
=\left(1-\frac{2M_G}{R}\right)
\exp\!\left[
-2\int_{P_s}^{P(r)}
\frac{d\widetilde P}{\varepsilon(\widetilde P)+\widetilde P}
\right]
\label{eq:C3}
\end{equation}
is normalized by the surface value \(e^{2\nu(R)}=1-2M_G/R\). 

Here \(\varepsilon\) and $P$ satisfy the EOS used in the TOV equations. The radius $R$ and total gravitational mass $M_G$ are determined by the pressure cutoff $P_s$. For the numerical accuracy consistent with the EOSs and solutions to the TOV equations, the adopted cutoff pressures \(P_s \) in S1, S2, and cold matter in units of \(\mathrm{MeV\,fm^{-3}}\) are \(2.52529\times10^{-13}\), \(1.56025\times10^{-14}\), and \(2.30854\times10^{-12}\), respectively.

\subsection{Radial pulsation equation and boundary conditions}\label{sec:appC2}

For a radial fluid element displacement \(\delta r(r,t)=u(r)e^{i\omega t}\), the linear radial pulsation equation takes the Sturm–Liouville form
\begin{equation}
\frac{d}{dr}\!\left(\Pi\frac{d\zeta}{dr}\right)
+\left(\mathcal{Q}+\omega^2W\right)\zeta=0,
\label{eq:C5}
\end{equation}
where the radial variable $\zeta(r)=r^2e^{-\nu(r)}u(r)$ and the coefficients
\begin{equation}
\begin{aligned}
\Pi&=\frac{\Gamma_{\mathrm{eq}} P}{r^2}e^{\lambda+3\nu},\quad W=\frac{\varepsilon+P}{r^2}e^{3\lambda+\nu}\\
\mathcal{Q}&=\frac{\varepsilon+P}{r^2}e^{\lambda+3\nu}
\left[
\left(\frac{d\nu}{dr}\right)^2
+\frac{4}{r}\frac{d\nu}{dr}
-8\pi P e^{2\lambda}
\right]
\end{aligned}
\label{eq:C6}
\end{equation}
are given by the background metric and EOSs.

The Lagrangian pressure perturbation $\Delta P$ is related to \(\zeta\) by
\begin{equation}
\Delta P =-\frac{\Gamma_{\mathrm{eq}} P}{r^2}e^\nu\frac{d\zeta}{dr},
\label{eq:C8}
\end{equation}
where the adiabatic index $\Gamma_{\mathrm{eq}}$ is given by the equilibrium EOSs and the corresponding speed of sound $c_{\mathrm{eq}}$
\begin{equation}
\Gamma_{\mathrm{eq}}=\frac{\varepsilon+P}{P}\,c_{\mathrm{eq}}^2,
\qquad
c_{\mathrm{eq}}^2
=\left.\frac{dP}{d\varepsilon}\right|_{\text{equilibrium EOS}}.
\label{eq:C7}
\end{equation}

\bibliographystyle{unsrt}
\bibliography{HQ_References_Combined}

@article{Pons:1998mm,
    author = "Pons, J. A. and Reddy, S. and Prakash, M. and Lattimer, J. M. and Miralles, J. A.",
    title = "{Evolution of protoneutron stars}",
    eprint = "astro-ph/9807040",
    archivePrefix = "arXiv",
    reportNumber = "SUNY-NTG-98-30",
    doi = "10.1086/306889",
    journal = "Astrophys. J.",
    volume = "513",
    pages = "780",
    year = "1999"
}

@article{Pons:2001ar,
    author = "Pons, Jose A. and Steiner, Andrew W. and Prakash, Madappa and Lattimer, James M.",
    title = "{Evolution of protoneutron stars with quarks}",
    eprint = "astro-ph/0102015",
    archivePrefix = "arXiv",
    doi = "10.1103/PhysRevLett.86.5223",
    journal = "Phys. Rev. Lett.",
    volume = "86",
    pages = "5223--5226",
    year = "2001"
}

@article{Yasutake:2009kj,
    author = "Yasutake, Nobutoshi and Kashiwa, Kouji",
    title = "{Lepton effects on the proto-neutron stars with the hadron-quark mixed phase in the Nambu-Jona-Lasinio model}",
    eprint = "0902.0111",
    archivePrefix = "arXiv",
    primaryClass = "astro-ph.HE",
    doi = "10.1103/PhysRevD.79.043012",
    journal = "Phys. Rev. D",
    volume = "79",
    pages = "043012",
    year = "2009"
}

@article{Shao:2011nu,
    author = "Shao, Guo-yun",
    title = "{Evolution of proto-neutron stars with the hadron-quark phase transition}",
    eprint = "1109.4340",
    archivePrefix = "arXiv",
    primaryClass = "nucl-th",
    doi = "10.1016/j.physletb.2011.09.030",
    journal = "Phys. Lett. B",
    volume = "704",
    pages = "343--346",
    year = "2011"
}

@article{Rikovska-Stone:2006gml,
    author = "Rikovska-Stone, J. and Guichon, Pierre A. M. and Matevosyan, Hrayr H. and Thomas, Anthony William",
    title = "{Cold uniform matter and neutron stars in the quark-mesons-coupling model}",
    eprint = "nucl-th/0611030",
    archivePrefix = "arXiv",
    reportNumber = "DAPNIA-06-321, JLAB-THY-06-546",
    doi = "10.1016/j.nuclphysa.2007.05.011",
    journal = "Nucl. Phys. A",
    volume = "792",
    pages = "341--369",
    year = "2007"
}

@article{Tiesinga:2021myr,
    author = "Tiesinga, Eite and Mohr, Peter J. and Newell, David B. and Taylor, Barry N.",
    title = "{CODATA recommended values of the fundamental physical constants: 2018}",
    doi = "10.1103/RevModPhys.93.025010",
    journal = "Rev. Mod. Phys.",
    volume = "93",
    number = "2",
    pages = "025010",
    year = "2021"
}

@article{Routaray:2023spb,
    author = "Routaray, Pinku and Mohanty, Sailesh Ranjan and Das, H. C. and Ghosh, Sayantan and Kalita, P. J. and Parmar, Vishal and Kumar, Bharat",
    title = "{Investigating dark matter-admixed neutron stars with NITR equation of state in light of PSR J0952-0607}",
    eprint = "2304.05100",
    archivePrefix = "arXiv",
    primaryClass = "nucl-th",
    doi = "10.1088/1475-7516/2023/10/073",
    journal = "JCAP",
    volume = "10",
    pages = "073",
    year = "2023"
}

@article{Ghosh:2023tbn,
    author = "Ghosh, Sayantan and Shaikh, Shahebaj and Kalita, Probit J. and Routaray, Pinku and Kumar, Bharat and Agrawal, B. K.",
    title = "{Exploring the macroscopic properties of proto-neutron stars: Effects of entropy and lepton fraction}",
    doi = "10.1016/j.nuclphysb.2024.116697",
    journal = "Nucl. Phys. B",
    volume = "1008",
    pages = "116697",
    year = "2024"
}

@article{Sagert:2008ka,
    author = "Sagert, I. and Fischer, T. and Hempel, M. and Pagliara, G. and Schaffner-Bielich, J. and Mezzacappa, A. and Thielemann, F. -K. and Liebendorfer, M.",
    title = "{Signals of the QCD phase transition in core-collapse supernovae}",
    eprint = "0809.4225",
    archivePrefix = "arXiv",
    primaryClass = "astro-ph",
    doi = "10.1103/PhysRevLett.102.081101",
    journal = "Phys. Rev. Lett.",
    volume = "102",
    pages = "081101",
    year = "2009"
}

@article{Hempel:2009vp,
    author = "Hempel, M. and Pagliara, G. and Schaffner-Bielich, J.",
    title = "{Conditions for Phase Equilibrium in Supernovae, Proto-Neutron and Neutron Stars}",
    eprint = "0907.2680",
    archivePrefix = "arXiv",
    primaryClass = "astro-ph.HE",
    doi = "10.1103/PhysRevD.80.125014",
    journal = "Phys. Rev. D",
    volume = "80",
    pages = "125014",
    year = "2009"
}

@article{Bombaci:2011mx,
    author = "Bombaci, Ignazio and Logoteta, Domenico and Providencia, Constanca and Vidana, Isaac",
    title = "{Effects of quark matter nucleation on the evolution of proto-neutron stars}",
    eprint = "1102.1665",
    archivePrefix = "arXiv",
    primaryClass = "astro-ph.SR",
    doi = "10.1051/0004-6361/201015783",
    journal = "Astron. Astrophys.",
    volume = "528",
    pages = "A71",
    year = "2011"
}

@article{Klahn:2016uce,
    author = "Klahn, Thomas and Fischer, Tobias and Hempel, Matthias",
    title = "{Simultaneous chiral symmetry restoration and deconfinement - Consequences for the QCD phase diagram}",
    eprint = "1603.03679",
    archivePrefix = "arXiv",
    primaryClass = "nucl-th",
    doi = "10.3847/1538-4357/836/1/89",
    journal = "Astrophys. J.",
    volume = "836",
    number = "1",
    pages = "89",
    year = "2017"
}

@article{Fischer:2017lag,
    author = {Fischer, Tobias and Bastian, Niels-Uwe F. and Wu, Meng-Ru and Baklanov, Petr and Sorokina, Elena and Blinnikov, Sergei and Typel, Stefan and Kl{\"a}hn, Thomas and Blaschke, David B.},
    title = "{Quark deconfinement as a supernova explosion engine for massive blue supergiant stars}",
    eprint = "1712.08788",
    archivePrefix = "arXiv",
    primaryClass = "astro-ph.HE",
    doi = "10.1038/s41550-018-0583-0",
    journal = "Nature Astron.",
    volume = "2",
    number = "12",
    pages = "980--986",
    year = "2018"
}

@article{Burgio:2010ek,
    author = "Burgio, G. F. and Schulze, H. -J.",
    title = "{The maximum and minimum mass of protoneutron stars in the Brueckner theory}",
    eprint = "1002.4497",
    archivePrefix = "arXiv",
    primaryClass = "astro-ph.SR",
    doi = "10.1051/0004-6361/201014308",
    journal = "Astron. Astrophys.",
    volume = "518",
    pages = "A17",
    year = "2010"
}

@article{Camelio:2017nka,
    author = "Camelio, Giovanni and Lovato, Alessandro and Gualtieri, Leonardo and Benhar, Omar and Pons, Jos{\'e} A. and Ferrari, Valeria",
    title = "{Evolution of a proto-neutron star with a nuclear many-body equation of state: Neutrino luminosity and gravitational wave frequencies}",
    eprint = "1704.01923",
    archivePrefix = "arXiv",
    primaryClass = "astro-ph.HE",
    doi = "10.1103/PhysRevD.96.043015",
    journal = "Phys. Rev. D",
    volume = "96",
    number = "4",
    pages = "043015",
    year = "2017"
}

@article{Raduta:2020fdn,
    author = "Raduta, Adriana R. and Oertel, Micaela and Sedrakian, Armen",
    title = "{Proto-neutron stars with heavy baryons and universal relations}",
    eprint = "2008.00213",
    archivePrefix = "arXiv",
    primaryClass = "nucl-th",
    doi = "10.1093/mnras/staa2491",
    journal = "Mon. Not. Roy. Astron. Soc.",
    volume = "499",
    number = "1",
    pages = "914--931",
    year = "2020"
}

@article{Issifu:2024fuw,
    author = "Issifu, Adamu and Menezes, D{\'e}bora P. and Rezaei, Zeinab and Frederico, Tobias",
    title = "{Proto-neutron stars with quark cores}",
    eprint = "2405.10386",
    archivePrefix = "arXiv",
    primaryClass = "nucl-th",
    doi = "10.1088/1475-7516/2025/01/024",
    journal = "JCAP",
    volume = "01",
    pages = "024",
    year = "2025"
}

@article{Laskos-Patkos:2026avp,
    author = "Laskos-Patkos, P. and Koliogiannis, P. S. and Moustakidis, C.",
    title = "{Hadron-quark phase transitions along proto-neutron-star evolution}",
    eprint = "2608.09195",
    archivePrefix = "arXiv",
    primaryClass = "nucl-th",
    month = "8",
    year = "2026",
    note = {arXiv:2608.09195}
}

@article{Lugones:1997gg,
    author = "Lugones, G. and Benvenuto, O. G.",
    title = "{Effect of trapped neutrinos in the hadron matter to quark matter transition}",
    doi = "10.1103/PhysRevD.58.083001",
    journal = "Phys. Rev. D",
    volume = "58",
    pages = "083001",
    year = "1998"
}

@article{Roark:2018uls,
    author = "Roark, J. and Dexheimer, V.",
    title = "{Deconfinement phase transition in proto-neutron-star matter}",
    eprint = "1803.02411",
    archivePrefix = "arXiv",
    primaryClass = "nucl-th",
    doi = "10.1103/PhysRevC.98.055805",
    journal = "Phys. Rev. C",
    volume = "98",
    number = "5",
    pages = "055805",
    year = "2018"
}

@article{Chandrasekhar:1964zz,
    author = "Chandrasekhar, S.",
    title = "{The Dynamical Instability of Gaseous Masses Approaching the Schwarzschild Limit in General Relativity}",
    doi = "10.1086/147938",
    journal = "Astrophys. J.",
    volume = "140",
    pages = "417--433",
    year = "1964",
    note = "[Erratum: Astrophys.J. 140, 1342 (1964)]"
}

@article{Gondek:1997fd,
    author = "Gondek, D. and Haensel, P. and Zdunik, J. L.",
    title = "{Radial pulsations and stability of protoneutron stars}",
    eprint = "astro-ph/9705157",
    archivePrefix = "arXiv",
    journal = "Astron. Astrophys.",
    volume = "325",
    pages = "217--227",
    year = "1997"
}

@article{Kokkotas:2000up,
    author = "Kokkotas, Kostas. D. and Ruoff, Johannes",
    title = "{Radial oscillations of relativistic stars}",
    eprint = "gr-qc/0011093",
    archivePrefix = "arXiv",
    doi = "10.1051/0004-6361:20000216",
    journal = "Astron. Astrophys.",
    volume = "366",
    pages = "565",
    year = "2001"
}

@article{Keil:1995hw,
    author = "Keil, W. and Janka, H. Th.",
    title = "{Hadronic phase transitions at supranuclear densities and the delayed collapse of newly formed neutron stars}",
    journal = "Astron. Astrophys.",
    volume = "296",
    pages = "145",
    year = "1995"
}

@article{Baumgarte:1996xz,
    author = "Baumgarte, T. W. and Shapiro, S. L. and Teukolsky, S. A.",
    title = "{Computing the delayed collapse of hot neutron stars to black holes}",
    doi = "10.1086/176849",
    journal = "Astrophys. J.",
    volume = "458",
    pages = "680",
    year = "1996"
}

@article{Fischer:2008rh,
    author = "Fischer, T. and Whitehouse, S. C. and Mezzacappa, A. and Thielemann, F. -K. and Liebendorfer, M.",
    title = "{The neutrino signal from protoneutron star accretion and black hole formation}",
    eprint = "0809.5129",
    archivePrefix = "arXiv",
    primaryClass = "astro-ph",
    doi = "10.1051/0004-6361/200811055",
    journal = "Astron. Astrophys.",
    volume = "499",
    pages = "1",
    year = "2009"
}

@article{Pereira:2017rmp,
    author = "Pereira, Jonas P. and Flores, C{\'e}sar V. and Lugones, Germ{\'a}n",
    title = "{Phase transition effects on the dynamical stability of hybrid neutron stars}",
    eprint = "1706.09371",
    archivePrefix = "arXiv",
    primaryClass = "gr-qc",
    doi = "10.3847/1538-4357/aabfbf",
    journal = "Astrophys. J.",
    volume = "860",
    number = "1",
    pages = "12",
    year = "2018"
}

@article{Lugones:2021zsg,
    author = "Lugones, Germ{\'a}n and Grunfeld, Ana Gabriela",
    title = "{Phase Conversions in Neutron Stars: Implications for Stellar Stability and Gravitational Wave Astrophysics}",
    doi = "10.3390/universe7120493",
    journal = "Universe",
    volume = "7",
    number = "12",
    pages = "493",
    year = "2021"
}

@article{Rather:2024hmo,
    author = "Rather, Ishfaq Ahmad and Marquez, Kauan D. and Backes, Betania C. and Panotopoulos, Grigoris and Lopes, Ilidio",
    title = "{Radial oscillations of hybrid stars and neutron stars including delta baryons: the effect of a slow quark phase transition}",
    eprint = "2401.07789",
    archivePrefix = "arXiv",
    primaryClass = "nucl-th",
    doi = "10.1088/1475-7516/2024/05/130",
    journal = "JCAP",
    volume = "05",
    pages = "130",
    year = "2024"
}

@article{Thakur:2025qwl,
    author = "Thakur, Prashant and Issifu, Adamu and Rather, Ishfaq Ahmad and Lim, Y. and Frederico, Tobias",
    title = "{Radial and non-radial oscillations of protoneutron stars with hyperonic composition}",
    eprint = "2505.24104",
    archivePrefix = "arXiv",
    primaryClass = "nucl-th",
    doi = "10.1140/epjc/s10052-026-16101-4",
    journal = "Eur. Phys. J. C",
    volume = "86",
    number = "7",
    pages = "863",
    year = "2026"
}

@article{Milthaler:2026dhm,
    author = "Milthaler, Jannik and Christian, Jan-Erik and Jakobus, Pia and Rosswog, Stephan",
    title = "{Twins in the Shadow: Phase Transitions in Proto-Neutron Stars}",
    eprint = "2610.03148",
    archivePrefix = "arXiv",
    primaryClass = "astro-ph.HE",
    month = "10",
    year = "2026",
    note = {arXiv:2610.03148}
}

@article{Rehberg:1995kh,
    author = "Rehberg, P. and Klevansky, S. P. and Hufner, J.",
    title = "{Hadronization in the SU(3) Nambu-Jona-Lasinio model}",
    eprint = "hep-ph/9506436",
    archivePrefix = "arXiv",
    reportNumber = "HD-TVP-95-06",
    doi = "10.1103/PhysRevC.53.410",
    journal = "Phys. Rev. C",
    volume = "53",
    pages = "410--429",
    year = "1996"
}

@article{Ratti:2005jh,
    author = "Ratti, Claudia and Thaler, Michael A. and Weise, Wolfram",
    title = "{Phases of QCD: Lattice thermodynamics and a field theoretical model}",
    eprint = "hep-ph/0506234",
    archivePrefix = "arXiv",
    doi = "10.1103/PhysRevD.73.014019",
    journal = "Phys. Rev. D",
    volume = "73",
    pages = "014019",
    year = "2006"
}

@article{Roessner:2006xn,
    author = "Roessner, Simon and Ratti, Claudia and Weise, W.",
    title = "{Polyakov loop, diquarks and the two-flavour phase diagram}",
    eprint = "hep-ph/0609281",
    archivePrefix = "arXiv",
    reportNumber = "ECT*-06-16",
    doi = "10.1103/PhysRevD.75.034007",
    journal = "Phys. Rev. D",
    volume = "75",
    pages = "034007",
    year = "2007"
}

@article{Sakai:2010rp,
    author = "Sakai, Yuji and Sasaki, Takahiro and Kouno, Hiroaki and Yahiro, Masanobu",
    title = "{Entanglement between deconfinement transition and chiral symmetry restoration}",
    eprint = "1006.3648",
    archivePrefix = "arXiv",
    primaryClass = "hep-ph",
    reportNumber = "SAGA-HE-257",
    doi = "10.1103/PhysRevD.82.076003",
    journal = "Phys. Rev. D",
    volume = "82",
    pages = "076003",
    year = "2010"
}

@article{Sasaki:2011wu,
    author = "Sasaki, Takahiro and Sakai, Yuji and Kouno, Hiroaki and Yahiro, Masanobu",
    title = "{Quark-mass dependence of the three-flavor QCD phase diagram at zero and imaginary chemical potential: Model prediction}",
    eprint = "1105.3959",
    archivePrefix = "arXiv",
    primaryClass = "hep-ph",
    doi = "10.1103/PhysRevD.84.091901",
    journal = "Phys. Rev. D",
    volume = "84",
    pages = "091901",
    year = "2011"
}

@article{Glendenning:1992vb,
    author = "Glendenning, Norman K.",
    title = "{First order phase transitions with more than one conserved charge: Consequences for neutron stars}",
    reportNumber = "LBL-30295-REV, LBL-30295",
    doi = "10.1103/PhysRevD.46.1274",
    journal = "Phys. Rev. D",
    volume = "46",
    pages = "1274--1287",
    year = "1992"
}

@article{Muller:1995ji,
    author = "Muller, Horst and Serot, Brian D.",
    title = "{Phase transitions in warm, asymmetric nuclear matter}",
    eprint = "nucl-th/9505013",
    archivePrefix = "arXiv",
    reportNumber = "IU-NTC-95-04",
    doi = "10.1103/PhysRevC.52.2072",
    journal = "Phys. Rev. C",
    volume = "52",
    pages = "2072--2091",
    year = "1995"
}

@article{Tolman:1939jz,
    author = "Tolman, Richard C.",
    title = "{Static solutions of Einstein's field equations for spheres of fluid}",
    doi = "10.1103/PhysRev.55.364",
    journal = "Phys. Rev.",
    volume = "55",
    pages = "364--373",
    year = "1939"
}

@article{Oppenheimer:1939ne,
    author = "Oppenheimer, J. R. and Volkoff, G. M.",
    title = "{On massive neutron cores}",
    doi = "10.1103/PhysRev.55.374",
    journal = "Phys. Rev.",
    volume = "55",
    pages = "374--381",
    year = "1939"
}

@article{Malfatti:2019tpg,
    author = "Malfatti, Germ{\'a}n and Orsaria, Milva G. and Contrera, Gustavo A. and Weber, Fridolin and Ranea-Sandoval, Ignacio F.",
    title = "{Hot quark matter and (proto-) neutron stars}",
    eprint = "1907.06597",
    archivePrefix = "arXiv",
    primaryClass = "nucl-th",
    doi = "10.1103/PhysRevC.100.015803",
    journal = "Phys. Rev. C",
    volume = "100",
    number = "1",
    pages = "015803",
    year = "2019"
}

@article{Ghaemmaghami:2025atk,
    author = "Ghaemmaghami, S. A. and Ghazanfari Mojarrad, M.",
    title = "{Hot hybrid neutron star matter within a statistical baryonic approach and the PNJL quark model}",
    doi = "10.1103/PhysRevD.111.023026",
    journal = "Phys. Rev. D",
    volume = "111",
    number = "2",
    pages = "023026",
    year = "2025"
}

@article{Sun:2024vdt,
    author = "Sun, T. T. and Zheng, Z. Y. and Chen, H. and Wei, J. B. and Burgio, G. F. and Schulze, H. -J.",
    title = "{Radial oscillations of protoneutron stars}",
    eprint = "2408.06599",
    archivePrefix = "arXiv",
    primaryClass = "nucl-th",
    doi = "10.1103/PhysRevD.111.043008",
    journal = "Phys. Rev. D",
    volume = "111",
    number = "4",
    pages = "043008",
    year = "2025"
}

@article{Prakash:1996xs,
    author = "Prakash, Madappa and Bombaci, Ignazio and Prakash, Manju and Ellis, Paul J. and Lattimer, James M. and Knorren, Roland",
    title = "{Composition and structure of protoneutron stars}",
    eprint = "nucl-th/9603042",
    archivePrefix = "arXiv",
    reportNumber = "SUNY-NTG-96-11, NUC-MINN-93-23-T",
    doi = "10.1016/S0370-1573(96)00023-3",
    journal = "Phys. Rept.",
    volume = "280",
    pages = "1--77",
    year = "1997"
}

@article{Rau:2026oxz,
    author = "Rau, Peter B. and Sabatucci, Andrea and Sedrakian, Armen",
    title = "{Radial Modes of Hot Neutrino-Trapped Hybrid Stars}",
    eprint = "2610.07241",
    archivePrefix = "arXiv",
    primaryClass = "astro-ph.HE",
    month = "10",
    year = "2026",
    note = {arXiv:2610.07241}
}

@article{Sabatucci:2026qcz,
    author = "Sabatucci, Andrea and Sedrakian, Armen",
    title = "{Isentropic Hybrid Stars in the Nambu-Jona-Lasinio Model: Effects of Neutrino Trapping}",
    eprint = "2603.19085",
    archivePrefix = "arXiv",
    primaryClass = "nucl-th",
    doi = "10.3390/particles9020061",
    journal = "Particles",
    volume = "9",
    number = "2",
    pages = "61",
    year = "2026"
}

@article{Fukushima:2008wg,
    author = "Fukushima, Kenji",
    title = "{Phase diagrams in the three-flavor Nambu-Jona-Lasinio model with the Polyakov loop}",
    eprint = "0803.3318",
    archivePrefix = "arXiv",
    primaryClass = "hep-ph",
    reportNumber = "YITP-08-19",
    doi = "10.1103/PhysRevD.77.114028",
    journal = "Phys. Rev. D",
    volume = "77",
    pages = "114028",
    year = "2008",
    note = "[Erratum: Phys.Rev.D 78, 039902 (2008)]"
}

@article{Ghosh:2024auj,
    author = "Ghosh, Sayantan and Zhao, Tianqi and Kumar, Bharat and Mohanty, Sailesh Ranjan",
    title = "{The role of adiabatic sound speeds in neutron star radial oscillations and stability}",
    eprint = "2401.08347",
    archivePrefix = "arXiv",
    primaryClass = "nucl-th",
    doi = "10.1088/1475-7516/2025/09/025",
    journal = "JCAP",
    volume = "09",
    pages = "025",
    year = "2025"
}

@article{Kunkel:2024otq,
    author = {Kunkel, Selina and Wystub, Stephan and Schaffner-Bielich, J{\"u}rgen},
    title = "{Determining the minimal mass of a proto-neutron star with chirally constrained nuclear equations~of state}",
    eprint = "2411.14930",
    archivePrefix = "arXiv",
    primaryClass = "nucl-th",
    doi = "10.1103/PhysRevC.111.035807",
    journal = "Phys. Rev. C",
    volume = "111",
    number = "3",
    pages = "035807",
    year = "2025"
}

@article{Salgado:2026xwx,
    author = "Salgado, Eduardo L. G. and Costa, Pedro and Provid{\^e}ncia, Constan{\c{c}}a",
    title = "{Isentropic thermodynamics across the hadron-quark mixed phase in a two-phase model with a PNJL quark description}",
    eprint = "2603.11081",
    archivePrefix = "arXiv",
    primaryClass = "hep-ph",
    doi = "10.1103/ltq6-ryt9",
    journal = "Phys. Rev. D",
    volume = "114",
    number = "1",
    pages = "014042",
    year = "2026"
}

@article{Wu:2025ilj,
    author = "Wu, Xuhao and Chu, Peng-Cheng and Ju, Min and Liu, He",
    title = "{Symmetry energy effect on hot nuclear matter and proto-neutron stars}",
    eprint = "2611.01726",
    archivePrefix = "arXiv",
    primaryClass = "nucl-th",
    doi = "10.1088/1674-1137/adaa58",
    journal = "Chin. Phys. C",
    volume = "49",
    number = "5",
    pages = "054102",
    year = "2025"
}

@article{Steiner:2000bi,
    author = "Steiner, A. and Prakash, M. and Lattimer, J. M.",
    title = "{Quark-hadron phase transitions in young and old neutron stars}",
    eprint = "nucl-th/0003066",
    archivePrefix = "arXiv",
    doi = "10.1016/S0370-2693(00)00780-2",
    journal = "Phys. Lett. B",
    volume = "486",
    pages = "239--248",
    year = "2000"
}

@article{Pagliara:2009dg,
    author = "Pagliara, G. and Hempel, M. and Schaffner-Bielich, J.",
    title = "{A new possible quark-hadron mixed phase in protoneutron stars}",
    eprint = "0907.3075",
    archivePrefix = "arXiv",
    primaryClass = "astro-ph.HE",
    doi = "10.1103/PhysRevLett.103.171102",
    journal = "Phys. Rev. Lett.",
    volume = "103",
    pages = "171102",
    year = "2009"
}

@article{Sun:2021cez,
    author = "Sun, Ting-Ting and Zheng, Zi-Yue and Chen, Huan and Burgio, G. Fiorella and Schulze, Hans-Josef",
    title = "{Equation of state and radial oscillations of neutron stars}",
    eprint = "2101.07515",
    archivePrefix = "arXiv",
    primaryClass = "nucl-th",
    doi = "10.1103/PhysRevD.103.103003",
    journal = "Phys. Rev. D",
    volume = "103",
    number = "10",
    pages = "103003",
    year = "2021"
}

@article{Burrows:1986me,
    author = "Burrows, Adam and Lattimer, James M.",
    title = "{The birth of neutron stars}",
    doi = "10.1086/164405",
    journal = "Astrophys. J.",
    volume = "307",
    pages = "178--196",
    year = "1986"
}

@article{Logoteta:2022hvb,
    author = "Logoteta, Domenico and Bombaci, Ignazio and Perego, Albino",
    title = "{Isoentropic equations of state of $\beta $-stable hadronic matter with a quark phase transition}",
    doi = "10.1140/epja/s10050-022-00708-8",
    journal = "Eur. Phys. J. A",
    volume = "58",
    number = "3",
    pages = "55",
    year = "2022"
}

@article{Annala:2017llu,
    author = "Annala, Eemeli and Gorda, Tyler and Kurkela, Aleksi and Vuorinen, Aleksi",
    title = "{Gravitational-wave constraints on the neutron-star-matter Equation of State}",
    eprint = "1711.02644",
    archivePrefix = "arXiv",
    primaryClass = "astro-ph.HE",
    reportNumber = "CERN-TH-2017-236",
    doi = "10.1103/PhysRevLett.120.172703",
    journal = "Phys. Rev. Lett.",
    volume = "120",
    number = "17",
    pages = "172703",
    year = "2018"
}

@article{Zhang:2022uin,
    author = "Zhang, Lin and Huang, Mei",
    title = "{Holographic cold dense matter constrained by neutron stars}",
    eprint = "2209.00766",
    archivePrefix = "arXiv",
    primaryClass = "nucl-th",
    doi = "10.1103/PhysRevD.106.096028",
    journal = "Phys. Rev. D",
    volume = "106",
    number = "9",
    pages = "096028",
    year = "2022"
}

@article{Riley:2019yda,
    author = "Riley, Thomas E. and others",
    title = "{A $NICER$ View of PSR J0030+0451: Millisecond Pulsar Parameter Estimation}",
    eprint = "1912.05702",
    archivePrefix = "arXiv",
    primaryClass = "astro-ph.HE",
    doi = "10.3847/2041-8213/ab481c",
    journal = "Astrophys. J. Lett.",
    volume = "887",
    number = "1",
    pages = "L21",
    year = "2019"
}

@article{Miller:2019cac,
    author = "Miller, M. C. and others",
    title = "{PSR J0030+0451 Mass and Radius from $NICER$ Data and Implications for the Properties of Neutron Star Matter}",
    eprint = "1912.05705",
    archivePrefix = "arXiv",
    primaryClass = "astro-ph.HE",
    doi = "10.3847/2041-8213/ab50c5",
    journal = "Astrophys. J. Lett.",
    volume = "887",
    number = "1",
    pages = "L24",
    year = "2019"
}

@article{Riley:2021pdl,
    author = "Riley, Thomas E. and others",
    title = "{A NICER View of the Massive Pulsar PSR J0740+6620 Informed by Radio Timing and XMM-Newton Spectroscopy}",
    eprint = "2105.06980",
    archivePrefix = "arXiv",
    primaryClass = "astro-ph.HE",
    doi = "10.3847/2041-8213/ac0a81",
    journal = "Astrophys. J. Lett.",
    volume = "918",
    number = "2",
    pages = "L27",
    year = "2021"
}

@article{Miller:2021qha,
    author = "Miller, M. C. and others",
    title = "{The Radius of PSR J0740+6620 from NICER and XMM-Newton Data}",
    eprint = "2105.06979",
    archivePrefix = "arXiv",
    primaryClass = "astro-ph.HE",
    doi = "10.3847/2041-8213/ac089b",
    journal = "Astrophys. J. Lett.",
    volume = "918",
    number = "2",
    pages = "L28",
    year = "2021"
}

@article{Breu:2016ufb,
    author = "Breu, Cosima and Rezzolla, Luciano",
    title = "{Maximum mass, moment of inertia and compactness of relativistic stars}",
    eprint = "1601.06083",
    archivePrefix = "arXiv",
    primaryClass = "gr-qc",
    doi = "10.1093/mnras/stw575",
    journal = "Mon. Not. Roy. Astron. Soc.",
    volume = "459",
    number = "1",
    pages = "646--656",
    year = "2016"
}

@article{Baumgarte:1999cq,
    author = "Baumgarte, Thomas W. and Shapiro, Stuart L. and Shibata, Masaru",
    title = "{On the maximum mass of differentially rotating neutron stars}",
    eprint = "astro-ph/9910565",
    archivePrefix = "arXiv",
    doi = "10.1086/312425",
    journal = "Astrophys. J. Lett.",
    volume = "528",
    pages = "L29",
    year = "2000"
}

\end{document}